\documentclass[twocolumn]{aastex63}
\pdfoutput=1
\usepackage{graphicx}
\graphicspath{{./}} 
\usepackage{advdate}

\renewcommand{\farcs}{\rlap{$^{\prime\prime}$}.\hbox to 2pt{}}
\renewcommand{\farcm}{\rlap{$^{\prime}$}.\hbox to 0.4198pt{}}

\newcommand{\msun}     {\ensuremath{{M}_{\scriptscriptstyle \odot}}}
\newcommand{\subsun}     {\ensuremath{_{\scriptscriptstyle \odot}}}
\newcommand{\kms}      {\ensuremath{\mathrm{km~s^{-1}}}}

\newcommand{\units}[1]  {\ensuremath{\mathrm{{#1}}}}

\received{\AdvanceDate[-1]\today}
\revised{\today}
\accepted{\AdvanceDate[1]\today}
\submitjournal{AAS Journals}

\shorttitle{\emph{Chandra} Observations of A2261-BCG}
\shortauthors{G{\"u}ltekin et al.}
\setwatermarkfontsize{108pt}

\begin{document}

\title{\textit{Chandra} Observations of Abell 2261 Brightest Cluster Galaxy,\\ a Candidate Host to a Recoiling Black Hole}

\correspondingauthor{Kayhan G{\"u}ltekin}
\email{kayhan@umich.edu}

\author[0000-0002-1146-0198]{Kayhan G{\"u}ltekin}
\affil{University of Michigan, Dept.\ of Astronomy, 1085 S.\ University Ave., Ann Arbor, MI, 48104, USA}

\author{Sarah Burke-Spolaor}
\affil{Department of Physics and Astronomy, West Virginia University, P.O. Box 6315, Morgantown, WV 26506, USA}
\affil{Center for Gravitational Waves and Cosmology, West Virginia University, Chestnut Ridge Research Building, Morgantown, WV 26505, USA}
\affil{Canadian Institute for Advanced Research, CIFAR Azrieli Global Scholar, MaRS Centre West Tower, 661 University Ave. Suite 505, Toronto ON M5G 1M1, Canada
}

\author{Tod R.\ Lauer}
\affil{NSF's National Optical Infrared Astronomy Research
Laboratory,\footnote{The NSF's NOIR Lab is operated by AURA, Inc
under cooperative agreement with NSF.}
P.O. Box 26732, Tucson, AZ 85726, USA}

\author{T.\ Joseph W.\ Lazio}
\affil{
Jet Propulsion Laboratory, California Institute of Technology, 4800 Oak Grove Blvd, M/S~67-201, Pasadena, CA  91109, USA}

\author[0000-0003-3030-2360]{Leonidas A.\ Moustakas}
\affil{Jet Propulsion Laboratory, California Institute of Technology, 4800 Oak Grove Blvd, M/S~169-506, Pasadena, CA, 91109, USA}

\author{Patrick Ogle}
\affil{Space Telescope Science Institute, 3700 San Martin Drive, Baltimore, MD 21218, USA}

\author[0000-0002-9365-7989]{Marc Postman}
\affil{Space Telescope Science Institute, 3700 San Martin Drive, Baltimore, MD 21218, USA}

\begin{abstract}
We use \emph{Chandra} X-ray observations to look for evidence of a recoiling black hole from the brightest cluster galaxy in  Abell 2261 (A2261-BCG).  A2261-BCG is a strong candidate for a recoiling black hole because of its large, flat stellar core, revealed by \emph{Hubble Space Telescope} imaging observations.  We took 100-ksec observations with \emph{Chandra} and combined it with 35 ksec of archival observations to look for low-level accretion onto a black hole of expected mass $M\sim10^{10}\ \msun$ that could possibly be located in one of four off-center stellar knots near the galaxy's center or else in the optical center of the galaxy or in the location of radio emission.  We found no X-ray emission arising from a point source in excess of the cluster gas and can place limits on the accretion of any black hole in the central region to a 2--7 keV flux below $4.3 \times 10^{-16}\ \units{erg\ s^{-1}\ cm^{-2}}$, corresponding to a bolometric Eddington fraction of about $10^{-6}$.  Thus there is either no $10^{10}\ \msun$ black hole in the core of A2261-BCG, or it is accreting at a low level.  We also discuss the morphology of the X-ray emitting gas in the cluster and how its asymmetry is consistent with a large dynamic event.
\end{abstract}


\section{Introduction} 
\label{sec:intro}
The galaxy cluster \object[ACO 2261]{Abell 2261}'s brightest cluster galaxy (\object[2MASX J17222717+3207571]{A2261-BCG} hereafter) is perhaps the best place to look for a recoiling or ejected central supermassive black hole.  Abell 2261 is a rich cluster with virial mass $M_{\mathrm{vir}} = (1.7 \pm 0.2) \times 10^{15}\ \msun\ h_{70}^{-1}$ within $R_{\mathrm{vir}} \approx 3\ h_{70}^{-1}$ Mpc \citep{2012ApJ...757...22C}.
A2261-BCG has a number of properties that set it apart from most other galaxies. First, it has one of the largest cores in its stellar surface brightness profile.  Early-type galaxies can be separated into two categories based on their surface brightness profiles: core and cusp \citep{1995AJ....110.2622L}.  Galaxies can have central brightness profiles that can be parameterized as a Nuker law,
\begin{equation}
    I(r) = 2^{(\beta - \gamma)/\alpha} I_{b} \left(\frac{r_{b}}{r}\right)^{\gamma} \left[1 + \left(\frac{r}{r_{b}}\right)^{\alpha}\right]^{(\gamma - \beta)/\alpha},
\end{equation}
where $I_b$ is the brightness at the core (or ``break'') radius, $r_b$,  $\gamma$ and $\beta$ are the power-law indices for $r \ll r_b$ and $r \gg r_b$, respectively,  and $\alpha$ is an index that sets the sharpness of the break \citep{1995AJ....110.2622L}.  For $\gamma > 0.5$, a galaxy is considered a power-law galaxy. For $\gamma < 0.3$, it is considered a core galaxy.  A2261-BCG has a core with $\gamma = -0.01$ and a cusp radius of $r_\gamma= 3.2\ \units{kpc}$ \citep{2012ApJ...756..159P}, where $r_\gamma$ is the  radius at which the local (negative) log-slope of the brightness profile is $1/2$.  Cores with {\it negative} values of $\gamma$ are seen in only rare cases \citep{laueretal02}, but imply that the central stellar density actually {\it decreases} as the center is approached. A2261-BCG thus has no central cusp in stellar density. A2261-BCG's core was the largest known at its time of discovery, and is still only smaller than those in Holm 15A, the brightest cluster galaxy in Abell 85 with $r_\gamma = 4.6\ \units{kpc}$ \citep{2014ApJ...795L..31L} and IC~1101, the brightest cluster galaxy in Abell~2029 with $r_b = 4.2\ \units{kpc}$ \citep{2017MNRAS.471.2321D}. Both of those galaxies, however, have shallow surface brightness cusps, which imply that their stellar density profiles are formally divergent at their centers, as is standard for nearly all core-galaxies.

In addition to having an unusually large and flat core, the stellar core of A2261-BCG is offset from the center of the galaxy as determined by the galaxy's outer isophotes \citep{2012ApJ...756..159P}.  The core's center is located $0.7$~kpc to the SW of the outer isophote center, suggestive of a recent perturbation.  Taken together, the extreme size and flatness of the core along with the offset, suggest that A2261-BCG is a possible site of a past black-hole--black-hole merger.

Black hole binaries are the leading candidate for creating core profiles in galaxy centers \citep{1980Natur.287..307B, 1997AJ....114.1771F}.  As two galaxies merge, their central black holes may sink to the center of the merged system from energy loss via dynamical friction \citep[e.g.,][but see also \citealt{2015MNRAS.451.1868T}]{1994MNRAS.271..317G, 2005ApJ...623L..67K}.  Eventually, enough energy is transferred from the black hole's orbits to the stars so as to gravitationally bind the two black holes in a binary system.  At this point, a major loss of energy from the binary black hole's orbit comes from interactions from stars passing by the binary.  As energy is transferred from the binary to the stars, the stars are flung out to large distances from the center of the galaxy.  This process turns what may have been a central surface brightness power-law into a core profile.  Thus a binary is said to have ``scoured'' a core into the galaxy.

If there are enough stars on orbits close enough to the binary, sufficient energy can be extracted such that gravitational wave emission from the binary black hole system can lead to 
the black holes merging in less than a Hubble time.  In the final stages of black hole coalescence, asymmetric emission of gravitational waves as a result of mass asymmetry or spin anisotropy can lead to recoil of the merged black hole at speeds of up to several thousand $\units{km\ s^{-1}}$ \citep{2007PhRvL..98w1101G, 2007PhRvL..98w1102C, 2011PhRvL.107w1102L}. 

The properties of the stellar surface-brightness of the core are also thought to result from the properties of the binary, such as mass and mass ratio.  Although determining precise values of the binary may be impossible \citep{2019MNRAS.484.2851L}, it is reasonable to expect that an exceptionally flat and large core would result from a massive binary with mass ratio ($q \equiv m_2 / m_1 \le 1$) close to unity.  One intriguing possibility is that the ejection of the merged black hole actually causes the core to ``rebound" due to the removal of the central mass, resulting in the core becoming even larger \citep{2004ApJ...607L...9M,2004ApJ...613L..37B,2008ApJ...678..780G}.

A number of different techniques all predict the A2261-BCG should host a very massive black hole in its center. Based on the 3.2 kpc size of its core, A2261-BCG would be expected to host a black hole with mass $M > 10^{10}\ \msun$ \citep{laueretal07}.  The $M$--$\sigma$ relation \citep{Kormendy2013} and the $M$--$L$ relation \citep{mcconnell2011} predict that the A2261-BCG black hole mass lies in the range $6 \times 10^{9}\ \msun < M < 1.1 \times 10^{10}\ \msun$. 

There are alternatives to binary core scouring to explain early-type galaxy cores, such as dissipationless collapse \citep{2006MNRAS.370..681N}, thermal evaporation of cold gas via AGN feedback \citep{2009AN....330..910N}, and the sinking of black holes post merger without binary formation \citep{2010ApJ...725.1707G}.  Despite these alternatives, core-scouring is generally considered the most likely, and thus the properties of A2261-BCG paint a picture in which a massive binary existed at one point.

Finding any evidence of current or previous binary black hole systems is salient for gravitational wave studies, especially low-frequency experiments such as pulsar timing arrays \citep[PTAs;][]{2013CQGra..30v4007H,  2019ApJ...880..116A, 2016MNRAS.458.3341D, 2016MNRAS.458.1267V} and the \emph{Laser Interferometer Space Antenna} \citep[\emph{LISA}][]{2017arXiv170200786A}.  The current electromagnetic evidence for unambiguous binary black holes is relatively sparse.  The strongest candidate is 0402+379, which has strong evidence based on VLBI observations with proper motion measurements \citep{2009ApJ...697...37R, 2017ApJ...843...14B}, but extensive searches for additional such objects have not produced more examples \citep{2011MNRAS.410.2113B}.  Other candidates exist but are less compelling as alternative explanations exist.  Cyg A was reported to contain a transient secondary source in infrared \citep{2003ApJ...597..823C} and radio \citep{2017ApJ...841..117P} observations, with the most likely interpretation being an infalling stripped galaxy with an accreting black hole, but the 460 pc projected separation may indicate that it is not gravitationally bound and is, instead, a dual AGN.
OJ287 has a light curve that may be explained as a binary system \citep{1988LNP...307...68V, 2011ApJ...729...33V, 2020ApJ...894L...1L} or a precessing BL~Lac jet \citep{2013MNRAS.434.2275T}.   A number of optical light curves from time-domain surveys such as Pan-STARRS \citep{2004AN....325..636H, 2016arXiv161205560C}, Catalina Real-Time Transient Survey \citep{2009ApJ...696..870D}, and Palomar Transient Factory \citep{2009PASP..121.1334R, 2009PASP..121.1395L} show variability that has been claimed to be periodic and thus evidence for a binary system \citep{2015ApJ...803L..16L,  2015Natur.518...74G, 2016MNRAS.463.2145C}, but these sources have also been explained as displaying typical AGN variability \citep{2016MNRAS.461.3145V} or having an SED consistent with a single AGN \citep{2017ApJ...851..106F}.  AGN with velocity-offset broad line regions are evidence for low-mass-ratio binaries \citep{2012ApJS..201...23E, 2015ApJS..221....7R, 2017MNRAS.468.1683R}, but the complexities of broad line regions make any interpretation difficult \citep{2012ApJS..201...23E}.  For the case of 0402+379, the projected separation is 7 pc, which puts it well out of the regime of gravitational wave emission and thus does not answer the question of whether black holes actually merge. 
Thus any evidence for black hole binaries is noteworthy, and evidence of black hole mergers is critical.

We have pursued evidence of a recoiling or ejected black hole in A2261-BCG as it would indicate that a binary black hole had existed and merged.  In particular, we have focused our past efforts on the core region of A2261-BCG, including targeted studies of 
four stellar knots\footnote{The knots have SIMBAD identifiers \object{[PLD2012] Knot 1}, \object{[PLD2012] Knot 2}, \object{[PLD2012] Knot 3}, and \object{[PLD2012] Knot 4}%
.}
near the center of the galaxy 
\citep{2012ApJ...756..159P}.  A recoiling black hole is expected to take with it a cloak of stars that are tightly bound to it.
One of the stellar knots could be the expected cloak of stars bound tightly to the recoiling black hole.  
We used observations of A2261-BCG with the Karl G. Jansky Very Large Array (VLA) to look for core radio emission coming from one of the stellar knots but found only evidence of old jet activity \citep{2017ApJ...849...59B}.  We also used \emph{HST}/STIS spectroscopy of the northern three stellar knots to look for stellar velocity dispersions that would indicate the presence of a massive object. The velocity dispersions measured in two of the knots were modest, and the third knot was poorly constrained because of low signal-to-noise ratio \citep{2017ApJ...849...59B}.  The fourth (southern) knot is too faint to attempt \emph{HST} spectroscopy in a reasonable amount of time.

In this paper we use deep \emph{Chandra} X-ray observations to look for evidence of accretion onto a large black hole at the core of A2261-BCG.  As discussed above, we expect the galaxy to at one time have hosted a $\sim10^{10}\,\msun$ black hole, and if the black hole has never suffered a recoil or if the recoil is modest (below $\sim750\ \kms$), then it will remain in the galaxy's core.  Accretion onto a black hole above an Eddington ratio of about $f_{\mathrm{Edd}} > 10^{-6}$ will be visible as an X-ray point source. We focus our search for such emission at six locations of interest: the locations of the four stellar knots, which may be hosting a recoiling supermassive black hole, the optical center of the BCG, which is the expected site of a non-recoiled black hole, and the location of radio emission found by \citet{2017ApJ...849...59B}, which is the location of previous AGN activity.

In Section \ref{sec:reduce}, we describe the \emph{Chandra} observations and our treatment of the data reduction, paying close attention to astrometric registration.  We present our analysis of the X-ray data in Section \ref{sec:analysis}, including image and spectral fitting, finding no evidence for an X-ray point source.  We discuss our results in Section \ref{sec:discuss} and suggest a potential way forward with high-spatial resolution infrared integral-field spectroscopy.  Throughout this paper, we assume a cosmology with $\Omega_m = 0.3$, $\Omega_\Lambda = 0.7$, and $H_0 = 70\ \units{km\ s^{-1}\ Mpc^{-1}}$.  The redshift of A2261-BCG is $z = 0.2233$, corresponding to a luminosity distance of $D_L = 1109\ \units{Mpc}$ and an angular diameter distance of $D_A = 741\ \units{Mpc}$, at which distance 1\arcsec\ subtends $3.59\ \units{kpc}$.

\section{Observations and Data Reduction}\label{sec:observe}
\label{sec:reduce}
Table~\ref{tab:log} summarizes the three \emph{Chandra} data sets that we use in our search for an X-ray signature of the central black hole in \hbox{A2261-BCG}.  The two archival data sets (ObsId~550 and~5007) were part of Guaranteed Time Observations (GTO) to measure the Sunyaev-Zeldovich effect and cluster gas properties \citep{2006ApJ...647...25B, 2007ApJ...666..835B, 2008ApJS..174..117M, 2011MNRAS.413..313H}.  Our new observations\footnote{%
Cycle~19, PI:~K.~G{\"u}ltekin, Proposal~19700260, approved for~100~ks in total
} were specifically designed to search for a low-luminosity accreting black hole in \hbox{A2261-BCG}.  For scheduling purposes, our program was split into two observations (ObsIDs~20413 and~21960).  Below we use the ObsID to identify an individual observation, if necessary.  The aim points for each observation are listed in Table\ \ref{tab:log}.

\begin{deluxetable*}{lcccccc}
\tablecaption{Observation Log\label{tab:log}}
\tablehead{%
\colhead{Obsid} & \colhead{MJD} & \colhead{Right Ascension} & \colhead{Declination} & \colhead{Exposure} & \colhead{Mode} & \colhead{Detector}\\[-1.5 ex]
\colhead{}          &             \colhead{}              & \multicolumn{2}{c}{(J2000)}                             & \colhead{(ks)} &  \colhead{} & \colhead{}
}
\startdata
550                & 51494 & $17^\mathrm{h}$~$22^\mathrm{m}$~27\,\fs60 & $+32\arcdeg$~07\arcmin~37\farcs20 & \phn9.06  & VFAINT & ACIS-I\\
5007               & 53018 & $17^\mathrm{h}$~$22^\mathrm{m}$~27\,\fs20 & $+32\arcdeg$~07\arcmin~58\farcs00 & 24.32 & VFAINT  & ACIS-I\\
20413\tablenotemark{a}              & 58441 & $17^\mathrm{h}$~$22^\mathrm{m}$~27\,\fs20 & $+32\arcdeg$~07\arcmin~57\farcs30 & 45.49 & \phm{V}FAINT  & ACIS-S\\
21960\tablenotemark{a}              & 58459 & $17^\mathrm{h}$~$22^\mathrm{m}$~27\,\fs20 & $+32\arcdeg$~07\arcmin~57\farcs30 & 49.44 & \phm{V}FAINT  & ACIS-S\\
\enddata
\tablenotetext{a}{Our data from the \emph{Chandra} program associated with proposal 19700260 were split into two separate exposures for a total of approximately 100 ks.}
\end{deluxetable*}

Our data reduction follows standard techniques with subpixel event repositioning techniques.  All data were re-reduced using the \emph{Chandra} Interactive Analysis of Observations (CIAO) v4.11 software package \citep{2006SPIE.6270E..1VF} and version 4.8.2 of the CALDB calibration database.  Subpixel events were achieved using the energy-dependent sub-pixel event repositioning algorithm  \citep[EDSER;][]{2004ApJ...610.1204L} from within the standard \texttt{chandra\_repro} reprocessing, which also generates the bad pixel files and level 2 event files.  We examined each observation individually for flares and removed intervals in which the count rate outside of obvious source areas was more than 3$\sigma$ from the mean count rate across each observation.
The new data show the cluster X-ray emission extends to 135\arcsec (725 kpc) with a total of $1.7 \times 10^{5}$ counts in this area.

\subsection{Astrometric Registration}

We pay careful attention to the absolute astrometric calibration of the \emph{Chandra} observations in order to look for X-ray sources at the locations of interest identified in \emph{HST}, Subaru/HSC, or VLA imaging.  Our strategy was to first match all \emph{Chandra} observations to each other, stack them, and then register the stacked \emph{Chandra} data to the Subaru data imaging, which has already been registered to the \emph{HST} imaging.

To register the X-ray data to each other we used standard \emph{Chandra} procedures to find point sources in each data set and register them all to a common world coordinate system. This was accomplished by creating a counts and image map and then making a point-spread function (PSF) map for each data set.  
We ran \texttt{wavdetect} on each image to detect point sources.  Then we used the CIAO tool \texttt{reproject\_aspect} to match point sources and perform the coordinate transform.  We used data set 21960 as the reference image to which the other three data sets were matched.  We tried various combinations of parameters but found that they made little difference in the end results and thus used the default settings.  In each case there were many common point sources allowing for a 4-parameter rotate--translate--scale transform.

Once we had the common coordinate system, we created new PSF maps for the new aspect solutions and then merged the four data sets using \texttt{merge\_obs} in the full band.  After this, we created an exposure-weighted mean PSF map of the merged data set.  We used this joint PSF map to run \texttt{wavdetect} on the merged image, finding 72 point sources.  We visually inspected each of these and rejected those that were low-quality.  Most of the point sources are expected to be background QSOs, though some of them are foreground stars.

We used the X-ray point sources to register the merged \emph{Chandra} data to a deep optical image of a $36\arcmin \times 36\arcmin$ region centered on A2261. These optical data were obtained with Suprime-Cam on the 8.2-meter Subaru telescope. Most of the details of the Subaru Suprime-Cam imaging data used here has been previously summarized \citep{2012ApJ...757...22C, Elinor2013}. In brief, the astrometrically-calibrated Subaru image used to align our X-ray and optical data was a co-added $V_J+R_c$ composite with a total integration time of 6.6 hours. The image is sampled at $0\farcs2$\ pixel$^{-1}$ with a final angular resolution of FWHM $\sim 0\farcs7$. The effective limiting magnitude of the co-added image is $R_c \approx 26.5$. We visually inspected the Subaru imaging at the coordinates of the X-ray point sources.  If there was not a single high-quality optical counterpart (including if there were multiple potential counterparts), then we dropped this source from our list.  We used the remaining 30 counterparts to register the transform using the CIAO tool \texttt{wcs\_match} for a rotate--translate--scale transform, which was a 0.78 sky pixel shift (plus a small amount of rotation and scale).  Finally, we used \texttt{wcs\_update} to update the aspect solution of the merged \emph{Chandra} data set.

\section{Analysis}
\label{sec:analysis}

We separate our analysis into three components.  We first present our analysis of the X-ray image and fits to the hot gas morphology in Section~\ref{sec:analysis:imagefit}.  We then analyze and fit the spectra of the irregular annuli recovered from the image fitting in Section~\ref{sec:analysis:spectral}.  Finally we look for evidence of X-ray point sources in excess of the cluster gas with spectral fits to the X-ray emission at the locations of interest using the annuli spectra as background in Section~\ref{sec:analysis:pointsource}.

\subsection{Image Fitting}
\label{sec:analysis:imagefit}

\begin{figure*}[ht!]
\centerline{\includegraphics[width=1.2\textwidth]{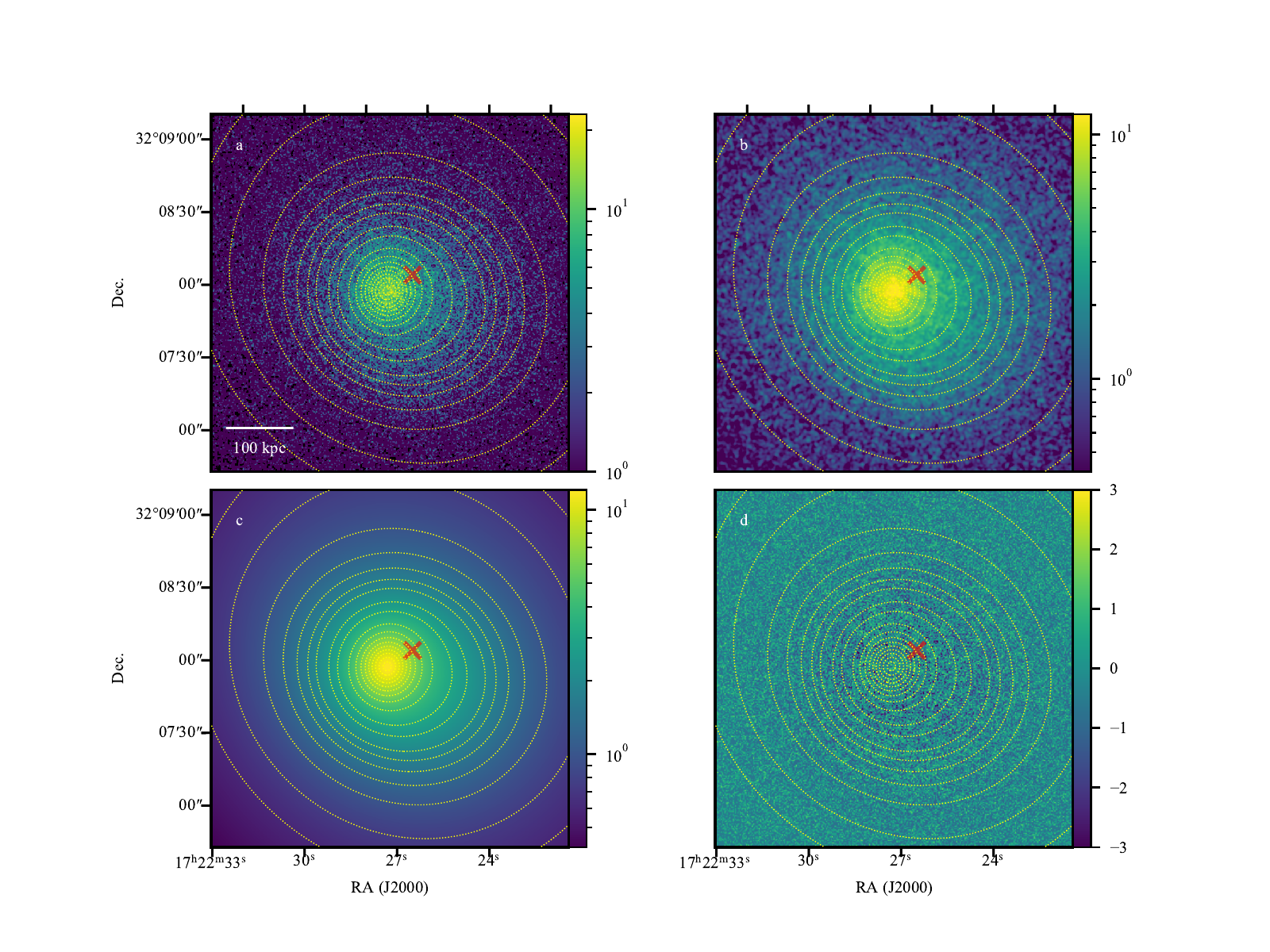}}
\caption{\emph{Chandra} 0.3--7~keV image and model of A2261.  The top panels show \emph{Chandra} data from the stacked image of the cluster A2261 with the left panel unsmoothed and binned at native \emph{Chandra} resolution ($0\farcs492\times0\farcs492$ pixels) and the right panel smoothed with a 5-pixel Gaussian kernel.  The bottom-left panel is the best-fit model to the data at the same colorscale as the smoothed data, but fit to the unsmoothed data.  The bottom-right panel shows residuals of the unsmoothed data from the best-fit model divided by the square root of the number of data counts (or by 1 if there are  no data counts).  
The bottom two panels show that the modeling is able to reproduce the major features of the data, including the elliptical isophotes and non-concentric inner/outer isophotes.  The $\times$ marks the location of the center of the outer-most isophotes and is clearly offset from the center of the inner isophotes.  We plot the isophotes of the best-fit model as dotted yellow lines in all panels. 
\label{fig:image}}
\end{figure*}

The combined X-ray image of A2261 shows a moderately disturbed morphology.   Figure~\ref{fig:image} shows that the X-ray emission is obviously non-circular, and examination of the isophotes shows that the eccentric isophotes are not concentric.  At scales larger than $\sim30\arcsec$, the eccentric isophotes are elongated along an axis roughly $40^\circ$ east of north with an eccentricity of $e \approx 0.5$.  At scales smaller than $\sim30\arcsec$, the isophotes are elongated along the  north--south direction with an eccentricity of $e \approx 0.4$.  The outer isophotes are centered at a location approximately $8\arcsec$ to the northwest of the inner isophote centers.

Using the aspect-corrected merged \emph{Chandra} data set, we performed two-dimensional fits to the imaging data with Sherpa \citep{2001SPIE.4477...76F}. 
To account for the non-circularity and non-concentricity, we fit the imaging data with a sum of three two-dimensional, eccentric beta models with independent central locations.  That is, each component has elliptical isophotes with eccentricity~$e$ and position angle $\theta$, centered at a position $x_0$, $y_0$, with a value that depends on $r$, the distance from the center, as $A [1 + (r / r_0)^2]^{-\alpha}$, where $A$ is the amplitude, $r_0$ is a core radius, and $\alpha$ is the index of the beta function.
The resulting fit (Fig.\ \ref{fig:image}c) matches the majority of the apparent visual properties of the data. 
In particular, the centers of the inner and outer (as determined by having small and large $r_0$, respectively) beta model components are separated by about $12\arcsec$ to the northwest and there is significant eccentricity in the outer component, aligned at $35^{\circ}$ east of north.  Similarly, we plot the residuals divided by the square root of the number of data counts (or by 1 when no counts) in Fig.\ \ref{fig:image}d.  The distribution of scaled residuals is approximately symmetric about zero, with a median of $-0.36$ and rms $0.74$.
The best-fit parameters and uncertainties are provided in Table \ref{tab:imagefit}.  

\begin{deluxetable}{lRRR}
\tablecaption{Imaging best fit}

\tablehead{\colhead{Parameter} & \colhead{Component 1} & \colhead{Component 2} & \colhead{Component 3}
 } 
\startdata
$r_0$    &        36_{-12}^{+3} &   431_{-168}^{+36}    &         548_{-96}^{+391} \\
$x_0$    & 4288.0_{-0.4}^{+0.6} & 4301.2_{-0.8}^{+4.2}  &     4308.5_{-1.8}^{+2.0} \\
$y_0$    & 4048.4_{-0.6}^{+0.5} & 4043.0_{-2.0}^{+0.7}  &     4061.7_{-1.8}^{+2.9} \\
$e$      & 0.09_{-0.04}^{+0.03} & 0.11_{-0.02}^{+0.03}  &            0.12 \pm 0.02 \\
$\theta$ & 4.5_{-0.2}^{+0.3}    & 2.61_{-0.08}^{+0.13}  &    -0.96_{-0.09}^{+0.07} \\
$A$     &    8.4_{-0.2}^{+1.2} &   2.6_{-0.6}^{+0.1}   &     0.99_{-0.07}^{+0.03} \\
$\alpha$ & 2.9_{-1.3}^{+0.9}    &       33_{-4}^{+16}   &        4.1_{-0.5}^{+2.9} \\
\enddata
\tablecomments{Parameters from the joint fit.  Locations and distances are measured in pixels, with the location $(4288.0,4048.4)$ corresponding to J2000 coordinates of 17:22:27.3, +32:07:57.54 with positive $x$ in the west direction and positive $y$ in the north direction.  The position angle~$\theta$ is in radians measured north of west, and the amplitude, $A$, is measured in units of normalized counts.}
\label{tab:imagefit}
\end{deluxetable}

\begin{figure*}[ht!]
\centerline{\includegraphics[width=1.2\textwidth]{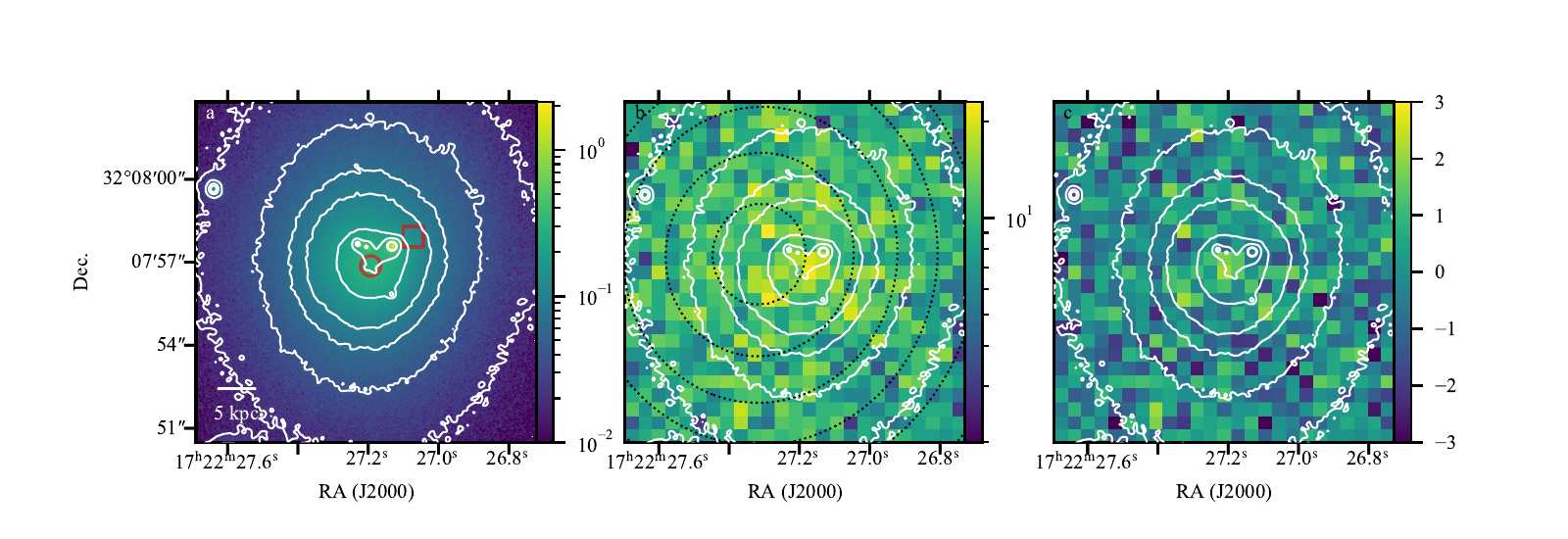}}
\caption{Images of the central region of \hbox{A2261~BCG}.  The panels show (a) the F606W \emph{HST}/ACS image from \citet{2012ApJ...756..159P} showing the 4 compact stellar knots within the central 3$\arcsec$, (b)~the stacked 0.3--7 keV \emph{Chandra} imaging data including our new 100~ks observations, and (c)~the residuals from our imaging fitting.  The X-ray data are binned to $0\farcs492\times0\farcs492$ pixels. The optical center (J2000 coordinates: 17:22:27.19, +32:07:56.87) and location of radio emission (J2000 coordinates: 17:22:27.07, +32:07:57.92) are shown in panel (a) as a red circle and square, respectively.   In all three panels, we show the contours of the F606W emission, which highlight the locations of the stellar knots.  In panel~b, we also show the contours of the best-fit X-ray image, which has inner isophotes centered at J2000 coordinates: 17:22:27.29, +32:07:57.55, offset from the stellar core.   There is no indication of excess emission indicative of an unmodeled X-ray point source at the locations of any of the four central stellar knots visible in the \emph{HST} image.
\label{fig:zoomimage}}
\end{figure*}

\subsection{Spectral Fitting}
\label{sec:analysis:spectral}
We fit spectral models of A2261 as well to determine if there is any excess hard emission at the location of the stellar knots, which would be indication of accretion onto a compact object.  Because there is significant X-ray emission from the cluster gas, it is necessary to model it.  First, we split the X-ray data into different annular regions; second, we fit the X-ray spectra to the projected annular regions; and third, we attempt to fit to the deprojected volume emission regions.  We detail each of these in turn.

Because of the nontrivial nature of the isophotes, we fit spectral models to annular regions as determined by isophote contours of the best-fit models.  The annuli are well approximated by elliptical boundaries, with varying center locations, eccentricity, and position angle.  The ellipses defining the annuli are presented in Table~\ref{tab:annuli} in the Appendix, along with the number of X-ray counts in each annulus. 

We did two X-ray spectral fits: first on each of the annular regions without deprojection, and then with an approximate deprojection of the three-dimensional spheroidal volumes that project to make the annular regions observed.  The advantage of the former approach is that it is straightforward while the advantage of the latter is that it allows inference of the underlying physical regions.  In both cases, we jointly fit each of the 4 observations, using their own individual response functions to properly account for the change in instrument behavior over time and the difference between ACIS-I and \hbox{ACIS-S}.
Fits were performed with XSPEC v12.10.1f using C-stat statistics.

For the projected case, each annular region is fitted with an intrinsically absorbed APEC model with a Galactic absorption value of $3.19 \times 10^{20}\ \units{cm^{-2}}$ and a redshift fixed at $z=0.2233$.   The spectral model we used produced good results at all annuli.  After doing an inital Levenberg-Marquett minimization to find best-fit parameters for each annulus, we computed the 68\% confidence intervals on each parameter with Markov chain Monte Carlo (MCMC) runs.  The best-fit and parameters and their uncertainties are presented in Table~\ref{tab:specfit} in the Appendix.

For the deprojected case, the entire cluster emission is fitted simultaneously using the XSPEC \texttt{projct} model to combine the volumes, which are determined by deprojecting the elliptical annuli into prolate ellipsoidal shells.  The deprojection requires that the elliptical annuli be concentric.  Despite the fact that the elliptical annuli are not concentric, we proceed as if they were all centered on the innermost region.  This assumption is unlikely to make a large difference as the change in center is small compared to the semi-minor axes of the larger ellipses (Table \ref{tab:annuli}), but it is formally incorrect.  We again use an absorbed APEC model with the same Galactic absorption and redshift as before.  The best-fit parameters and their 68\% confidence intervals were determined with MCMC runs and again presented in Table \ref{tab:specfit} in the Appendix.

\begin{figure*}[ht!]
\includegraphics[width=0.49\textwidth]{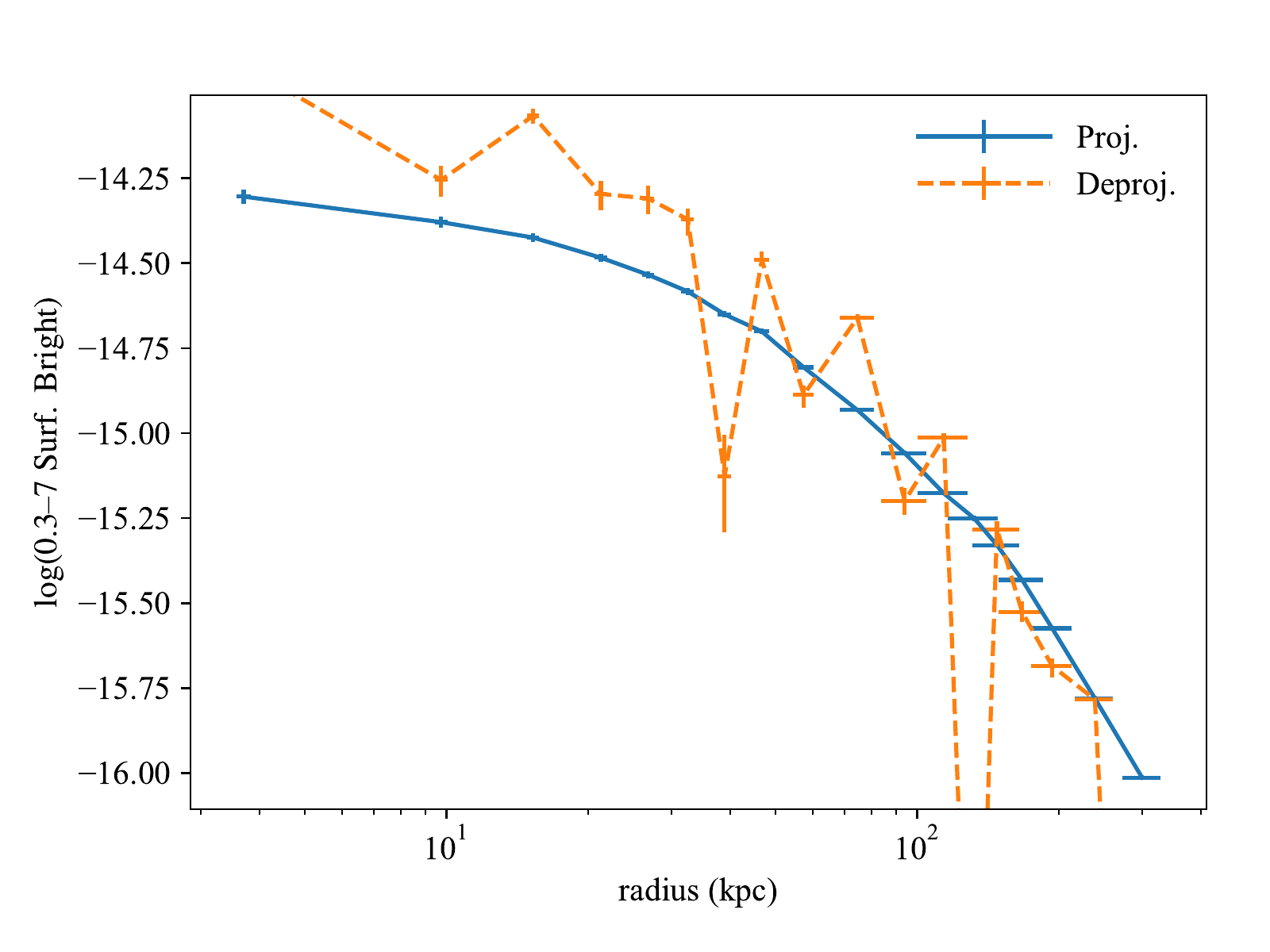}
\includegraphics[width=0.49\textwidth]{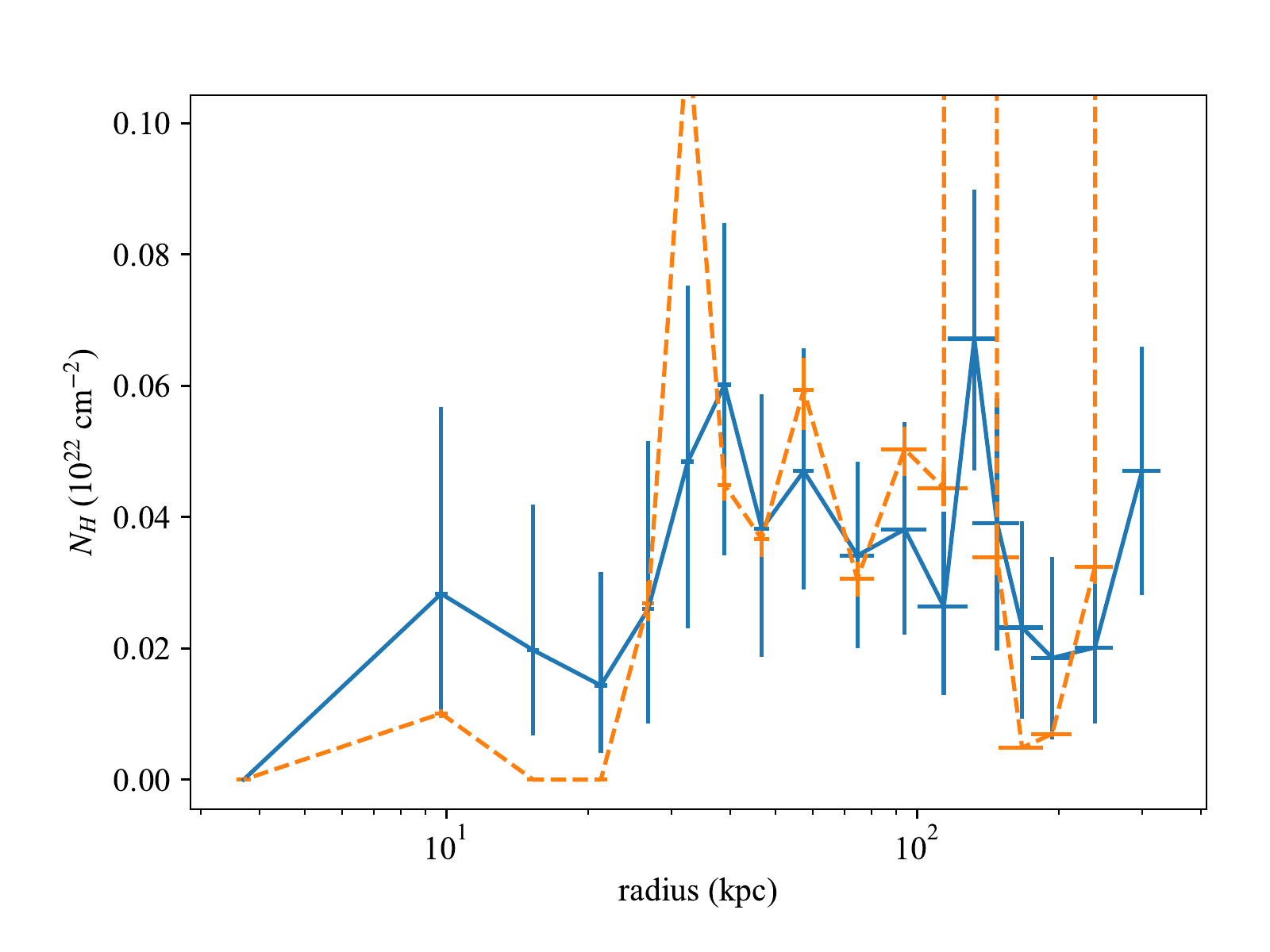}\\
\includegraphics[width=0.49\textwidth]{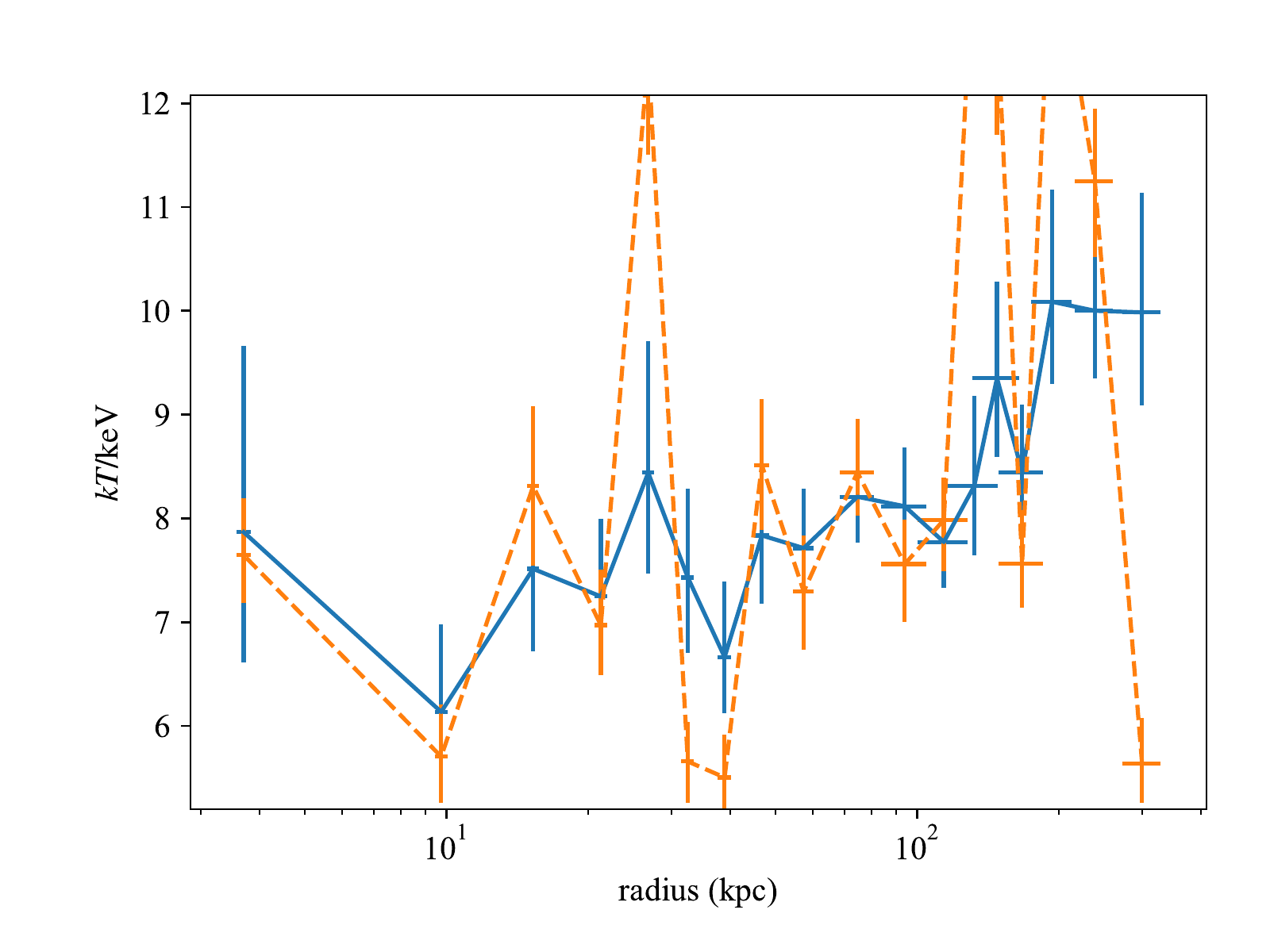}
\includegraphics[width=0.49\textwidth]{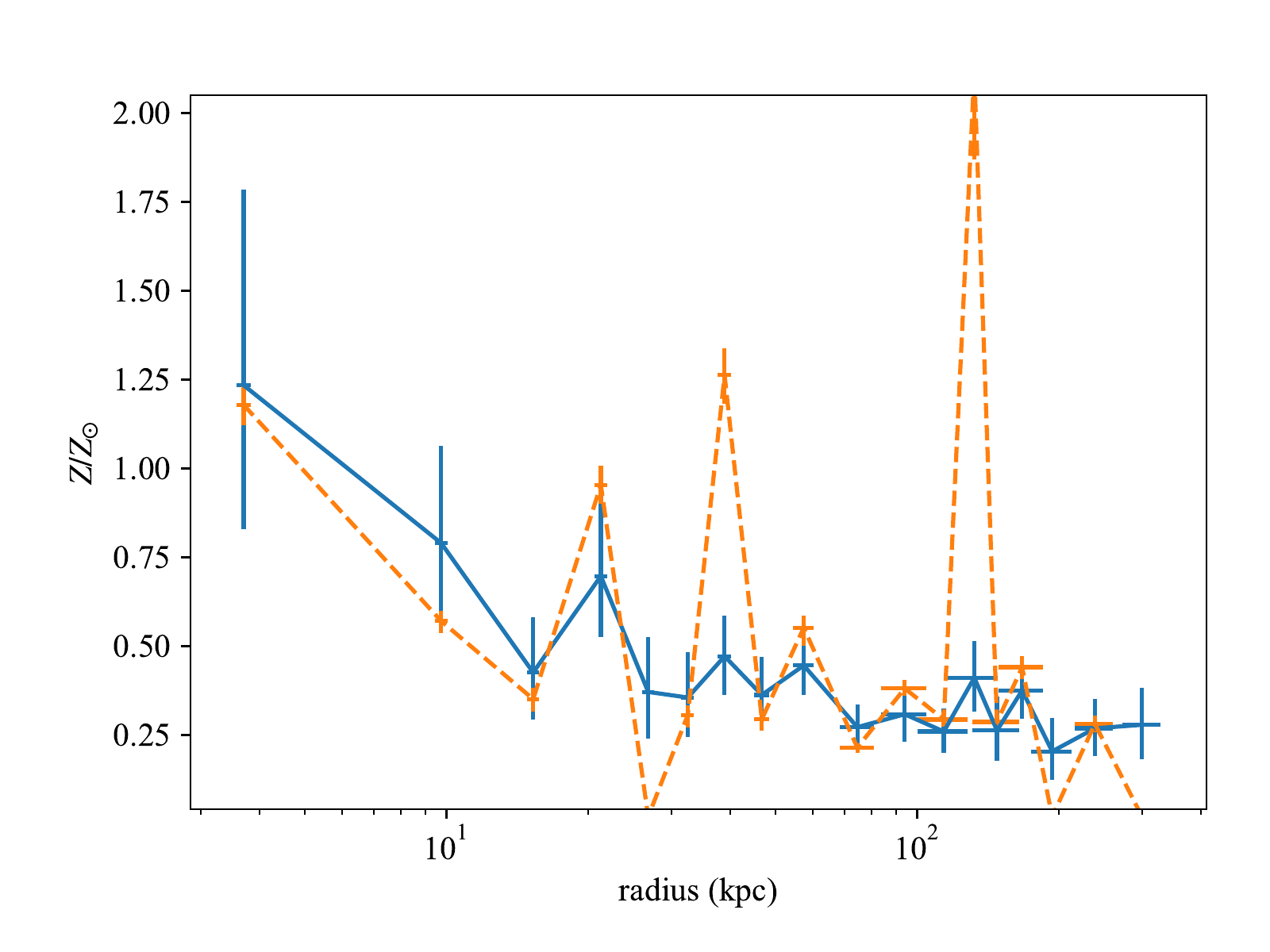}\\

\caption{Radial profiles of parameter estimates from spectral fits.  The 0.3--7 keV surface brightness is in units of $\units{erg\ s^{-1}\ cm^{-2}\ arcsec^{-2}}$, and the other panels are in the units as indicated. The solid blue line is from projected spectral fits, and the dashed orange line is from the deprojected spectral fits.  The radius of each annulus is determined by averaging the average radial distance to the inner and outer contours used to bound each region.  The horizontal error bars indicate the standard deviation in the radial distances.  The smoothness of the surface brightness profile is a result of the fact that the annular regions were based on the isophotes.  There is general agreement in the trends between the two different sets of fits.  The deprojected fits appear to have  variance larger than the reported statistical uncertainties.  The larger variance is likely the result of unaccounted systematic uncertainty arising from using a concentric model on non-concentric data.
\label{fig:profiles}}
\end{figure*}

\begin{figure}[htb]
\includegraphics[width=\columnwidth]{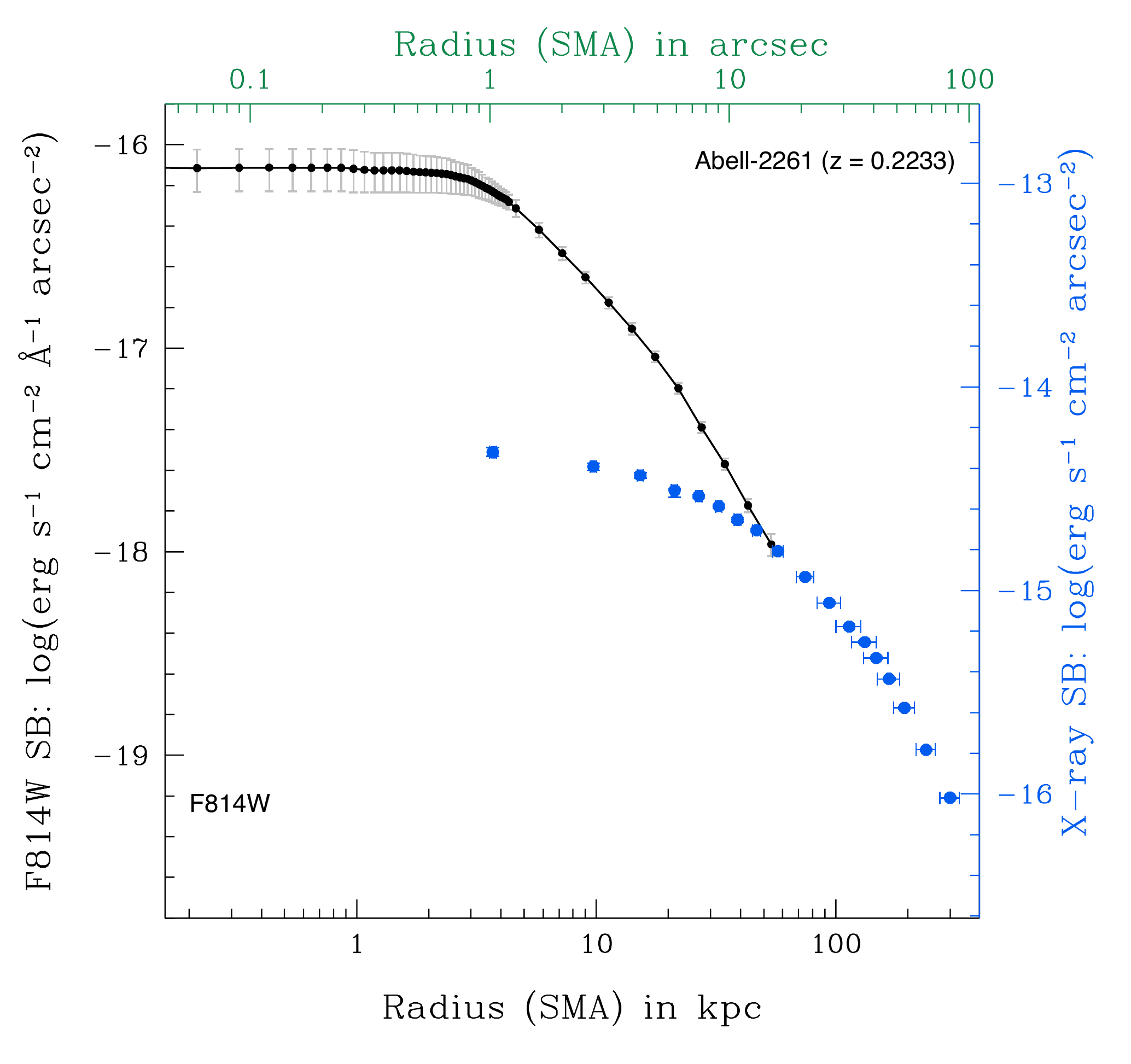}
\caption{Comparison of stellar light profile to the X-ray light profile. The black points that extend further to the left make up the F814W surface brightness profile from \emph{HST} data originally presented in \citet{2012ApJ...756..159P} and follow the vertical axis on the left.  The blue points that extend to the right are the same points from the X-ray surface brightness profile in the top-left panel of Fig.\ \ref{fig:profiles} and follow the vertical axis on the right.  The flatness and size of the stellar core are apparent, and the X-ray surface brightness profile reaches to angular sizes comparable to stellar brightness core.
}
\label{fig:profilecompare}
\end{figure}

We  plot profiles of the parameters in Fig.\ \ref{fig:profiles}.  Although there are differences between the projected fits and deprojected fits, they show the same general trends.  
The profiles show a few interesting trends across radius. First, the surface brightness profile of the projected case is smooth because the regions were selected based on the isophotes.  We compare the X-ray surface brightness profile to the stellar light profile in Fig.\ \ref{fig:profilecompare}.  The intrinsic absorption column shows a potential change from higher values at radii $r > 40\ \mathrm{kpc}$ to values consistent with no intrinsic absorbing column, though with significant uncertainties in the inner region because of the lower number of counts arising from the fact that the area is smaller.  In any event, the overall absorption is modest.  The temperature of the collisional plasma shows a steady increase from the inner regions of 6 to 9 keV at 200 kpc.  Finally the metallicity shows significant enhancement inside of 10 kpc compared to the rest of the cluster gas.

\begin{deluxetable}{lRRR}
\tablecaption{X-ray Point Source Upper Limits}

\tablehead{\colhead{Location} & \colhead{$x$} & \colhead{$y$} &  \colhead{99\% Limit}\\
\colhead{} & \colhead{} &\colhead{} &\colhead{\units{erg\ s^{-1}\ cm^{-2}}}
}
\startdata
Knot 1 & 1.41 &  0.05 & 2.9 \times 10^{-16} \\
Knot 2 & 2.50 &  0.00 & 2.7 \times 10^{-16} \\
Knot 3 & 3.96 &  0.05 & 2.7 \times 10^{-16} \\
Knot 4 & 3.94 & -3.46 & 4.3 \times 10^{-16} \\
Optical center & 2.61 & -1.37 & 2.3 \times 10^{-16} \\
Radio emission & 5.74 & 0.77  & 2.0 \times 10^{-16}
\enddata
\tablecomments{Upper limits to X-ray emission from the locations listed.  The locations are identified by name and with their coordinates in units of 0\farcs492 pixels relative to (4288.0,4048.4), which corresponds to J2000 coordinates of 17:22:27.3, +32:07:57.54.  The radii of the circular regions used in the X-ray analysis were all 1.04 pixels, except for knot 4, which was 1.25 pixels.   The last column lists the 99\% confidence upper limit to the restframe 2--7 keV flux, assuming a $\Gamma = 2$ power-law source with only Galactic absorption.}
\label{tab:limits}
\end{deluxetable}

\subsection{X-ray Point-Sources}
\label{sec:analysis:pointsource}
Our analysis shows no point-source X-ray emission arising from the location of the stellar knots, the optical center of the galaxy, or the radio emission.
First, when subtracting the best-fit model from the data, the residuals show no obvious excess emission at any of the above locations (Fig.\ \ref{fig:zoomimage}).  We quantify any potential excess by extracting emission from the locations of interest and fitting their spectra with a simple absorbed power-law model.  For each location, we consider the flux from a circular region centered on the knot with radius $0\farcs615$ (except for knot 4, which has radius $0\farcs739$) using the annular region in which it lies as the background region (with any other potential source locations excluded).  The positions of the regions are given in Table \ref{tab:limits}.  We then calculate flux upper limits assuming a $\Gamma = 2$ power-law spectral form with Galactic absorption ($N_H = 3.34 \times 10^{20}\ \units{cm^{-2}}$) for the rest-frame 2--7~keV band.  We fit all four observations simultaneously assuming the same spectral parameters.  All best-fit spectral models have fluxes indicating no emission above the background.  To calculate an upper limit to the flux, we used the XSpec \texttt{steppar} routine to find the value of flux at which the change in fit statistic was 6.635 for a 99\% confidence upper limit.  We list these limits in Table \ref{tab:limits}.  The highest of any of these limits is an unabsorbed flux of $4.3 \times 10^{-16}\ \mathrm{erg\ s^{-1}\ cm^{-2}}$.  Assuming a bolometric correction of 10 \citep{2007MNRAS.381.1235V}, this corresponds to a bolometric luminosity of $L = 6.3 \times 10^{41}\ \units{erg \ s^{-1}}$, which corresponds to an Eddington fraction limit of $f_{\mathrm{Edd}} < 5 \times 10^{-7} \approx 10^{-6}$, assuming a $M = 10^{10}\ \msun$ black hole.

\section{Discussion}
\label{sec:discuss}
Either there is no $10^{10}\ \msun$ black hole in any of the four stellar knots, the optical center, or the site of radio emission; or any such black hole is accreting below an Eddington fraction of $10^{-6}$.  We find no excess X-ray emission attributable to a point source in any of the four stellar knots near the galaxy core center, which for two of the knots is consistent with our past spectroscopic findings of low-mass ($<6\times10^8~\msun$) black holes \citep{2017ApJ...849...59B}. It is, of course, possible to find excess emission that could be attributed to a point source at many locations in the image, but, without a specific location to search, the chance of a spurious signal is high.  Thus, we restrict our search for excess emission to only the locations of interest.  As mentioned in \S\ref{sec:analysis:imagefit}, we are able to place limits on any accretion arising from the locations of interest.  Given the size of \hbox{A2261-BCG}, a black hole with mass following either the $M$--$\sigma$ or $M$--$L$ relations would be near $10^{10}\ \msun$.  Assuming a bolometric correction of 10 for hard X-ray, our flux limits correspond to emission at $10^{-6}$ of Eddington for a $10^{10}\ \msun$ black hole.  Given the range of bolometric corrections for hard X-rays at low accretion rates and the uncertainties associated with estimating black hole masses in such extreme environments, this Eddington fraction limit can only be considered an order-of-magnitude estimate.

The fact that we find an upper limit to the Eddington fraction of $f < 10^{-6}$ does not preclude a black hole from existing in any of the locations we searched.  
Eddington fractions below $10^{-6}$ are plausible for systems such as \hbox{A2261-BCG}.  Most BCGs do not show bright X-ray AGN at their centers \citep{2009ApJ...704.1586S, 2020ApJ...889..121D}.  
There has been considerable research into low-luminosity AGN.  By necessity, this research tends to focus on nearby sources such as Sgr A* \citep{2003ApJ...591..891B}, analogs of which cannot be detected outside of the Local Group.  Similarly, \citet{2006ApJ...640..126S} studied several elliptical galaxies with black holes having masses smaller than $6 \times 10^{8}\ \msun$ emitting in X-rays at fractions of Eddington below $\log(L_X / L_{\mathrm{Edd}}) = -6.96$.  Their sample includes \object[NGC 821]{NGC 821}, which was studied in detail by \citet{2007ApJ...667..731P} to have a 0.3--8 keV nuclear luminosity of $2.8 \times 10^{38}\ \units{erg\ s^{-1}}$.
M87 is the nearest central dominant galaxy in a large galaxy cluster and has a large black hole \citep[$M = 6.6 \times 10^{9}\ \msun$,][]{gebhardtetalm87} emitting at a small fraction of Eddington \citep{2003ApJ...582..133D}.  It is possible to infer the presence of a black hole from analysis of the cluster X-ray emission, but this requires well resolving the central kpc region, which is not possible for A2261 \citep{2018MNRAS.477.3583R}.

For comparison, \citet{2012ApJ...749..129G} and \citet{2019ApJ...871...80G} looked at nearly all galaxies with direct, dynamical black hole mass measurements as part of a broader study of the fundamental plane of black hole accretion.  Within their sample of SMBHs, they found that $L_X / L_{\mathrm{Edd}}$ ranged from $10^{-10}$ to $10^{-2}$, with roughly 60\% below $10^{-6}$.  The sample of galaxies is biased towards low accretion rates and quiescent AGN by the nature of needing to avoid a bright AGN when measuring stellar velocity distributions at the center.  If A2261-BCG were closer and/or had a tip-tilt star, it would similarly be a target for black hole mass estimation and could therefore be considered as broadly coming from the same parent distribution of galaxies.  On the other hand, the sample contained few BCGs, which may have qualitatively different fueling environments than other SMBH hosts, and there are not enough known $10^{10}\ \msun$ black holes to draw broad conclusions.  

High-resolution X-ray observations have shown that BCGs in general are unlikely to have a nuclear X-ray point source, despite other compelling evidence of an accreting black hole.  In particular, \citet{2011MNRAS.413..313H} looked at sample of BCGs (including A2261-BCG) in cool-core clusters with X-ray luminosities above $L_X = 10^{45}\ \units{erg\ s^{-1}}$. Their sample had radio emission, indicating the BCG hosted an accreting black hole at least at some time in the past.  None of the objects in their sample showed nuclear X-ray point sources.  \citet{2011MNRAS.413..313H} identified several potential reasons for the lack of X-ray nuclei in galaxies with obvious AGN radio jets: (i) very massive black holes (i.e., $M > 10^{10}\ \msun$) resulting in radiative power below the kinetic power inferred from the radio jets, (ii) X-ray emission originating in the jet and therefore Doppler suppressed out of the line of sight, (iii) alternative accretion mechanisms such as magnetically dominated accretion disks \citep{1982MNRAS.199..883B}, (iv) a large column absorbing the X-ray emission of the AGN, and (v) AGN duty cycles that indicate radio jet emission coming from a different epoch than that probed by X-ray observations.
Our VLA observations of A2261-BCG show that the radio emission with luminosity density $L_{\nu} \sim 5 \times 10^{30}\ \units{erg\ s^{-1}\ Hz^{-1}}$ is indeed from an earlier epoch (at least 48 Myr old based on synchrotron aging) and that it is thus not necessarily a strong predictor of X-ray emission \citep{2017ApJ...849...59B}.  That is, the radio emission is not evidence of a currently accreting black hole in A2261-BCG.  The questions raised about obscuration are, however, still important, especially as A2261-BCG has been identified as a potentially centrally obscured source.  We address these questions below.

We have determined that there is not significant obscuration preventing us from seeing black hole emission.  Here, we describe the arguments in the literature advocating for a significant dust mass and cold gas at the center of A2261-BCG and our line of reasoning for why A2261-BCG is not heavily obscured.  \citet{2008MNRAS.389.1637B} noted that its combined X-ray and optical properties were similar to those of blue-core, star-forming BCGs.  They argued that because A2261-BCG had a red core, this indicated that there was significant star formation that was extincted and that this was consistent with an earlier measurement of $\sim10^{8}\ \msun$ of dust in the galaxy.  This earlier measurement is due to \citet{2002MNRAS.330...92C} based on James Clerk Maxwell Telescope Submillimeter Common User Bolometer Array (JCMT/SCUBA) detection of significant emission at 450 and 850 $\mu$m.  This emission, however, comes from a location that is about 35\arcsec\ to the West of \hbox{A2261-BCG}.  The SCUBA beamsize is $\sim15\arcsec$ at 850 $\mu$m with JCMT pointing accuracty of about $2\arcsec$ \citep{2002MNRAS.330...92C}, which puts the 850 $\mu$m emission close to a region of 100 and 160 $\mu$m emission as detected by \emph{Herschel}/PACS \citep{herschel2019}.  
Further, we do not find evidence of significant ongoing or recent star formation in \hbox{A2261-BCG}.   First, \emph{Herschel}/PACS 100 and 160 $\mu$m imaging  shows no significant emission from A2261-BCG \citep{herschel2019}. Second, the SDSS spectrum of A2261-BCG is devoid of H$\alpha$, \ion{O}{3}, or other star-formation tracing lines, nor does the spectrum show evidence of strong Balmer absorption features.  Without any positive observational evidence for recent star formation, the red core of A2261-BCG is not sufficient reasoning to argue in favor of a heavy absorption.  Additionally, our analysis of optical BCG position relative to the X-ray peak gets a value of about $R_{\mathrm{off}} = 5.7 h_{70}^{-1}\ \units{kpc}$ compared to the value of  $R_{\mathrm{off}} = 0.4 h_{70}^{-1}\ \units{kpc}$ due to \citet{2008MNRAS.389.1637B}.  Adopting our larger value puts A2261-BCG much closer to the rest of the normal red-core BCGs.

\emph{HST}-resolution imaging also allows us to better see the inner stellar distribution and lack of dust lanes. A second line of reasoning by \citet{2008MNRAS.389.1637B} is the appearance of a double nucleus in their optical imaging as evidence for a dust lane.  Their optical imaging was done with MegaCam on the Canada-France-Hawaii Telescope with 0\farcs186 pixels in 0\farcs7 seeing.  It is likely that the stellar knots seen in \emph{HST} imaging appeared as a double nucleus. At higher resolution, the optical and infrared imaging show no evidence of dust, either.  The \emph{HST} UV, optical, and NIR imaging shows no central dust lanes or color gradients associated with dust obscuration \citep{2012ApJ...756..159P}.  Additionally, an estimated 3$\sigma$ upper limit to the 100 $\mu$m flux within the central 15 kpc is 0.24 mJy using the public Herschel science archive data products. The 0.2--2.0 $\mu$m spectral energy distribution of A2261-BCG also is well fit by models that have intrinsic extinction of $A_V \le 0.35\ \mathrm{mag}$. All of these constraints suggest that we have a largely dust-free view of the center of A2261-BCG.

We also find no evidence for large amounts of cold gas in A2261-BCG.  A final argument for heavy obscuration comes from the \emph{Spitzer} 3.6--24 $\mu$m observations \citep{2006ApJ...647..922E}.  The observations show extended emission with monochromatic luminosities of $\sim10^{42}$--$10^{44}\ \units{erg\ s^{-1}}$.  \citet{2011MNRAS.413..313H} argue that such infrared emission is to be expected as reprocessed emission by gas surrounding an AGN.
The amount of emission seen in the \emph{Spizter} imaging, however, is consistent with extrapolations of the stellar population SED fitting with only a slight excess at 24 $\mu$m.  Thus, the infrared emission is almost entirely starlight, indicating no evidence for excess cold gas in the system.  
%
Thus we conclude that we have a clear view of any accreting black hole, but no X-ray point source emission is seen.  It is still possible, however, that a $10^{10}\ \msun$ black hole exists in one of the stellar knots of A2261-BCG but is accreting at such a low level that it does not produce sufficient X-ray emission to be noticeable above the cluster gas emission.

Although absence of evidence is not evidence of absence, our observations are consistent with no $10^{10}\ \msun$ black hole existing in any of locations of the stellar knots.  If there is no black hole in any of the stellar knots, then the questions regarding the location of the expected central black hole of A2261-BCG remain.  One possibility is that a $10^{10}\ \msun$ black hole exists at the local center of the galaxy and is accreting at a very low rate such that there is no obvious X-ray point source.  Just as we cannot definitively ascertain whether there is a black hole in the stellar knots, we cannot definitely ascertain whether one exists at the center of the galaxy, either; but nor can we rule it out with the X-ray data.  
Another possibility is that the black hole did suffer a gravitational wave recoil of relatively high velocity ($\sim1000\ \units{km\ s^{-1}}$) over 10~Myr ago and it has traveled over 10 kpc from the center.  Looking at a wider radius of stellar knots suspected of being a recoiling black hole cloak would increase the chance of X-ray noise leading to an X-ray point source false positive.

The findings here are consistent in light of our past results, which as previously mentioned demonstrated through spectroscopy and stellar color analysis that two of the central stellar knots are likely to be low-mass group-member galaxies that are being stripped and cannibalized by the BCG. The existence of a relic radio source implies that there was at some point a sufficiently high accretion rate onto the black hole to support the formation of collimated jets, however the present lack of a bright X-ray core supports the conclusion that the radio source seen in this BCG represents relic emission.

It should be possible to definitively determine whether a black hole exists at the center of the A2261-BCG with \emph{JWST} IFU observations and full Schwarzschild modeling of the galaxy.  The central surface brightness of A2261-BCG is too low for \emph{HST} to measure the stellar velocity dispersion in the central $0.25\arcsec$ region, and there are no tip-tilt stars available for ground-based adaptive optics IFS observations.  With \emph{JWST} it would be possible to get high $S/N$ spectroscopy of the center of the galaxy to get stellar kinematics as well as to do so for the stellar knots.  Although the stellar knots had spectra taken with \emph{HST}/STIS, the hot pixels on STIS led to lower than expected $S/N$ and may have compromised the ability to measure the stellar velocity dispersions of the knots.  If a large black hole is present in any of the stellar knots, then it should be obvious in \emph{JWST} spectroscopy.  Similarly, if a $10^{10}\ \msun$ black hole is either present or absent from the center of the galaxy, it would be straightforward to detect its presence with Schwarzschild modeling of the same data.  If a large black hole were quickly removed from the center of the galaxy, the stars in the previous sphere of influence would react on a dynamical time.

\begin{figure}[htb]
\includegraphics[width=1.0\columnwidth]{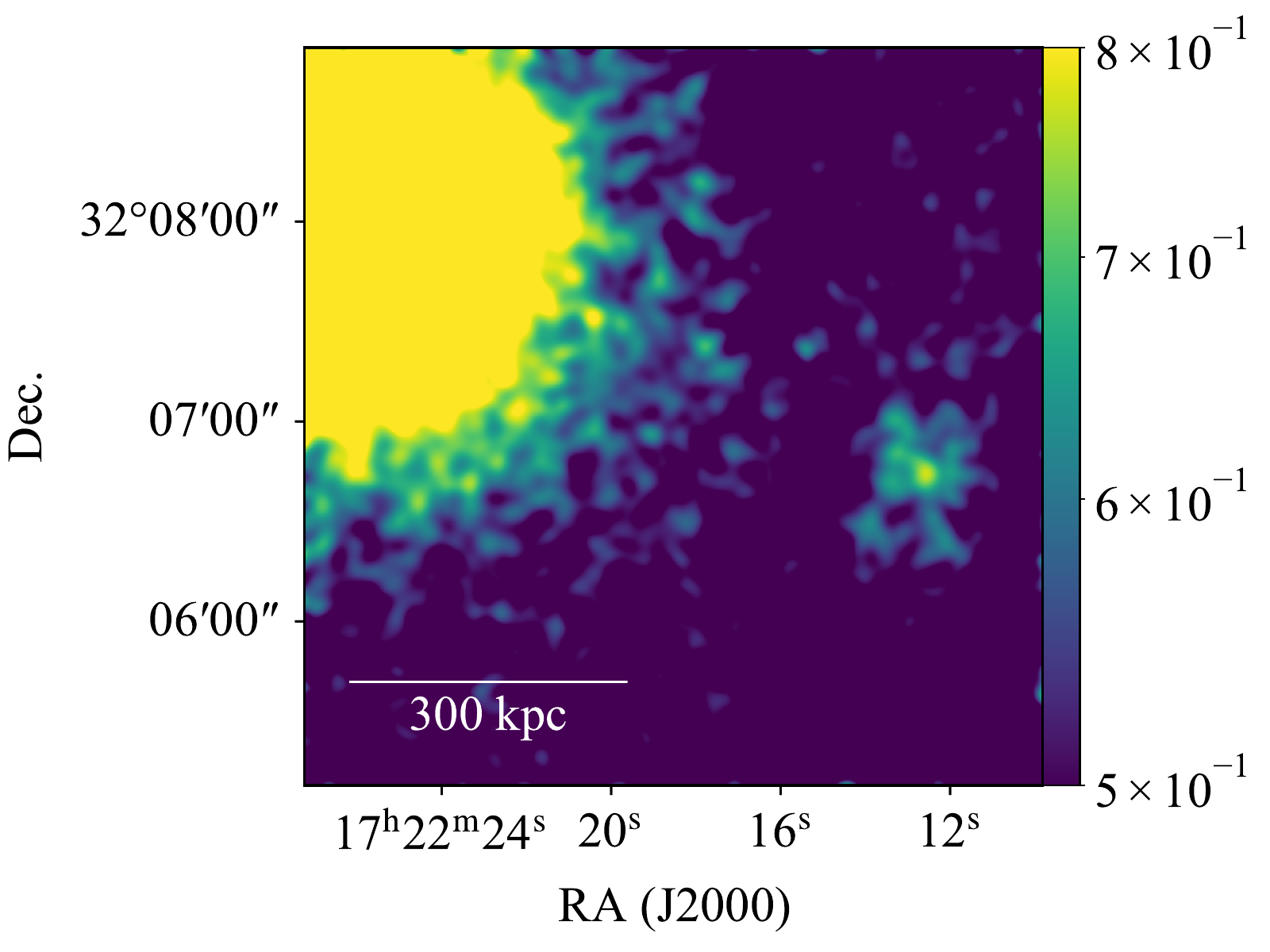}
\caption{Full-band X-ray image to the Southwest of the cluster center.  The color scale is set to show the cluster substructure, which is evident at the right-hand side of the figure.  The image has been smoothed with a Gaussian kernel with a standard deviation of 5 pixels.
}
\label{fig:zoomwayoutimage}
\end{figure}

We also note that the asymmetry in the X-ray morphology indicates a disturbance that is similar to the optical morphology disturbance.  Both the X-ray and optical light have cores that are displaced with respect to the centers as determined by outer isophotes.  The optical light, which traces starlight, and the X-ray light, which traces the hot gas, are similar in their amplitude and direction of displacement.  The X-ray morphology could be evidence of ``sloshing.'' Sloshing, the presence of asymmetric and/or spiral structure patterns in the hot gas in a cluster observed in some cool-core clusters \citep{2000ApJ...541..542M, 2003ASPC..301...37M,  2004ApJ...616..178C, 2010A&A...516A..32G, 2010MNRAS.405...91S, 2020A&A...633A..42S}, is thought to be a result of dark-matter substructure passing through the center of the cluster \citep{2006ApJ...650..102A, 2010ApJ...717..908Z,  2012MNRAS.420.3632R}.
The X-ray emission located 3\farcm3 southwest of the cluster core center may be evidence of this (Fig.\ \ref{fig:zoomwayoutimage}).  Located at $\alpha=17$:22:12.5, $\delta=+32$:06:43.4, there is extended emission in a roughly elliptical region extending $30\arcsec \times 20\arcsec$.  At a projected separation of approximately 700 kpc, this region contains several galaxies that are cluster members as determined by their redshift.  Thus the X-ray and optical morphology\,---\,including the large and offset stellar core\,---\,could have both been an effect of cluster substructure interactions.

\section{Summary}
In this work, we presented new 100~ks of \emph{Chandra} observations of \hbox{A2261-BCG}.  We fitted these new data along with 35~ks of archival data to determine the profile of X-ray emitting soft gas.  Our analysis of the morphology of the X-ray gas showed evidence of a previous dynamical disturbance, consistent with the optical morphology.  We also fitted the X-ray spectra of the extended emission, which yielded profiles showing a cool core with enhanced metallicity.  Using our cluster gas profile as a background region, we fitted the X-ray spectra at the locations of the stellar knots, the optical center, and the location of the radio emission.  Our analysis found that all of the locations of interest were consistent with no X-ray point sources.  Synthesizing available information about A2261-BCG, while this galaxy should host a SMBH with a mass potentially in excess of $10^{10}\ \msun$ and contains multiple indications that it could have been the site of a past merger of two SMBHs, we find no evidence for such a SMBH accreting above an Eddington fraction of $f_{\mathrm{Edd}} = 10^{-6}$.

\acknowledgments
KG acknowledges support provided by the National 
Aeronautics and Space Administration (NASA) through \emph{Chandra} Award Numbers GO7-18087X, TM8-19007X, and GO8-19078X issued by the \emph{Chandra} X-ray Observatory Center, which is operated by the Smithsonian Astrophysical observatory for and on behalf of NASA under contract NAS8-03060. 
SBS acknowledges support for this work by NSF grant \#1815664.
SBS is a CIFAR Azrieli Global Scholar in the Gravity and the Extreme Universe program.
The NANOGrav project receives support from National Science Foundation (NSF) Physics Frontier Center award number 1430284.
Part of this research was carried out at the Jet Propulsion Laboratory, California Institute of Technology, under a contract with the National Aeronautics and Space Administration.
The results in this paper are based, in part, on data collected at Subaru Telescope, which is operated by the National Astronomical Observatory of Japan.
This research has made use of NASA's Astrophysics Data System. This research used data products from the \emph{Herschel} Science Archive. \emph{Herschel} is an ESA space observatory with science instruments provided by European-led Principal Investigator consortia and with important participation from NASA.

\software{CIAO \citep{2006SPIE.6270E..1VF}, Sherpa \citep{2001SPIE.4477...76F, 2007ASPC..376..543D}, XSPEC \citep{1996ASPC..101...17A}, matplotlib \citep{2007CSE.....9...90H}, astropy \citep{2013A&A...558A..33A, 2018AJ....156..123A}, xvista \citep{1988igbo.conf..443S}}

\facilities{\emph{HST}, \emph{Chandra}, Subaru, \emph{Herschel}}

\clearpage
\bibliography{gultekin.bib}

\appendix
\label{sec:appendix}
In this appendix we present plots of the spectra for each annulus from which we derive the parameter profiles in Fig.\ \ref{fig:profiles}.

\begin{deluxetable}{cRRRRRR}
\tablecaption{Elliptical annuli}

\tablehead{\colhead{Annulus} & \colhead{Center $x$} & \colhead{Center $y$} & \colhead{PA} & \colhead{$a$} & \colhead{$b$}
 & \colhead{Counts}} 
\startdata
1 & 0.4 & -0.1 & 164.5 & 4.3 & 4.0 & 552\\
2 & 0.4 & -0.1 & 169.5 & 7.1 & 6.5 & 765\\
3 & 0.5 & -0.1 & 169.7 & 10.8 & 9.9 & 1544\\
4 & 0.5 & -0.1 & 170.0 & 14.0 & 12.9 & 1617\\
5 & 0.7 & -0.2 & 170.3 & 17.3 & 15.9 & 1807\\
6 & 0.8 & -0.2 & 170.9 & 20.7 & 19.2 & 2060\\
7 & 1.1 & -0.3 & 172.0 & 24.6 & 22.9 & 2473\\
8 & 1.7 & -0.5 & 174.8 & 29.5 & 27.8 & 3396\\
9 & 3.1 & -1.0 & 7.0 & 36.3 & 35.0 & 4805\\
10 & 6.7 & -2.2 & 49.6 & 49.0 & 46.5 & 8264 \\
11 & 9.3 & -3.1 & 55.7 & 59.5 & 54.9 & 6263\\
12 & 11.6 & -3.6 & 56.8 & 73.0 & 66.5 & 8273\\
13 & 12.4 & -3.6 & 56.7 & 81.4 & 73.9 & 5248\\
14 & 13.2 & -3.3 & 56.2 & 91.5 & 82.8 & 6320\\
15 & 13.9 & -2.7 & 55.1 & 104.4 & 94.3 & 8002\\
16 & 14.9 & -0.9 & 52.1 & 122.9 & 111.1 & 10809\\
17 & 17.3 & 5.5 & 41.8 & 157.4 & 142.3 & 19678\\
18 & 19.5 & 11.5 & 36.0 & 197.4 & 175.5 & 22121\\
\enddata
\tablecomments{Ellipses determining the annuli of regions used in spectral fitting.  Region $n$ is formed with ellipse in row $n$ as the outer boundary and row $n-1$ as the inner boundary, with the innermost region as an ellipse, rather than an annulus.  The center locations are in units of 0\farcs492 pixels relative to $(4288.0, 4048.4)$, which corresponds to J2000 coordinates of 17:22:27.3, +32:07:57.54. Position angle (PA) is given in units of degrees east of north, and the semimajor and semiminor axes are given in units of pixels.  The final column lists the number of X-ray counts in each annulus in the new observations.
The drift in ellipse centers demonstrates the non-concentricity of the ellipses.  Background regions used for X-ray spectral fitting were four circles with $(x, y, r)$ values of $(578.3, 299.0, 130.4)$, $(578.3, -202.3, 126.5)$, $(43.4, -339.3, 119.6)$, and $(33.0, 363.6, 96.3)$ again relative to the same origin as above.}
\label{tab:annuli}
\end{deluxetable}

\movetabledown=5.5cm
\begin{rotatetable*}
\begin{deluxetable*}{cRRRRRRRRR}
\tablecaption{Spectral fits}

\tablehead{\colhead{Annulus} & \multicolumn{2}{c}{$N_{\mathrm{H}}$} &  \multicolumn{2}{c}{$\log{F_X}$} & \multicolumn{2}{c}{$kT$} & \multicolumn{2}{c}{$Z / Z\subsun$} & $C$\mathrm{\mbox{-}stat}\\
&
\colhead{Proj.} & \colhead{Deproj.} &
\colhead{Proj.} & \colhead{Deproj.} &
\colhead{Proj.} & \colhead{Deproj.} &
\colhead{Proj.} & \colhead{Deproj.}
} 
\startdata
1 & 9.5^{+2194.9}_{-9.3} \times 10^{-8} & (8.4 \pm 0.7) \times 10^{-9} & -13.20 \pm 0.02 & -12.82^{+0.04}_{-0.03} & 7.87^{+1.80}_{-1.25} & 7.65^{+0.55}_{-0.46} & 1.23^{+0.55}_{-0.40} & 1.18 \pm 0.06 & 1300.0 \\
2 & 2.8^{+2.8}_{-1.9} \times 10^{-2} & (1.0 \pm 0.1) \times 10^{-2} & -13.04 \pm 0.02 & -12.91^{+0.04}_{-0.05} & 6.14^{+0.84}_{-0.65} & 5.70^{+0.50}_{-0.44} & 0.79^{+0.27}_{-0.23} & 0.57 \pm 0.03 & 1400.6  \\
3 & 2.0^{+2.2}_{-1.3} \times 10^{-2} & (1.4 \pm 0.1) \times 10^{-6} & -12.76 \pm 0.01 & -12.40 \pm 0.02 & 7.51^{+0.84}_{-0.79} & 8.31 \pm 0.77 & 0.43^{+0.16}_{-0.13} & 0.35 \pm 0.03 & 1692.4  \\
4 & 1.4^{+1.7}_{-1.0} \times 10^{-2} & 3.3^{+0.1}_{-0.2} \times 10^{-9} & -12.73 \pm 0.01 & -12.54^{+0.04}_{-0.05} & 7.25^{+0.74}_{-0.66} & 6.97^{+0.54}_{-0.48} & 0.70^{+0.21}_{-0.17} & 0.95 \pm 0.05 & 1710.1  \\
5 & 2.6^{+2.6}_{-1.7} \times 10^{-2} & (2.7 \pm 0.3) \times 10^{-2} & -12.68 \pm 0.01 & -12.46^{+0.04}_{-0.05} & 8.44^{+1.26}_{-0.97} & 12.45^{+0.85}_{-0.95} & 0.37^{+0.15}_{-0.13} & 0.020 \pm 0.001 & 1757.4  \\
6 & 4.8^{+2.7}_{-2.5} \times 10^{-2} & (1.1 \pm 0.1) \times 10^{-1} & -12.62 \pm 0.01 & -12.41^{+0.03}_{-0.05} & 7.43^{+0.86}_{-0.73} & 5.66^{+0.38}_{-0.40} & 0.36^{+0.13}_{-0.11} & 0.31^{+0.03}_{-0.04} & 1719.3  \\
7 & 6.0^{+2.5}_{-2.6} \times 10^{-2} & 4.5^{+0.3}_{-0.2} \times 10^{-2} & -12.544 \pm 0.010 & -13.02^{+0.12}_{-0.16} & 6.66^{+0.73}_{-0.54} & 5.50^{+0.41}_{-0.35} & 0.47^{+0.12}_{-0.11} & 1.26 \pm 0.08 & 1803.7  \\
8 & 3.8^{+2.1}_{-2.0} \times 10^{-2} & 3.7^{+0.2}_{-0.3} \times 10^{-2} & -12.413^{+0.008}_{-0.009} & -12.20 \pm 0.02 & 7.84^{+0.69}_{-0.66} & 8.51^{+0.64}_{-0.62} & 0.36^{+0.11}_{-0.09} & 0.29^{+0.02}_{-0.03} & 1873.2  \\
9 & 4.7^{+1.9}_{-1.8} \times 10^{-2} & 5.9^{+0.5}_{-0.6} \times 10^{-2} & -12.271 \pm 0.007 & -12.35^{+0.03}_{-0.04} & 7.71^{+0.57}_{-0.53} & 7.30^{+0.54}_{-0.56} & 0.45^{+0.09}_{-0.08} & 0.55^{+0.03}_{-0.05} & 2043.0  \\
10 & (3.4 \pm 1.4) \times 10^{-2} & (3.1 \pm 0.3) \times 10^{-2} & -12.047 \pm 0.006 & -11.78 \pm 0.01 & 8.21^{+0.51}_{-0.44} & 8.44^{+0.52}_{-0.42} & 0.27 \pm 0.06 & 0.21 \pm 0.01 & 2100.6  \\
11 & (3.8 \pm 1.6) \times 10^{-2} & 5.0^{+0.3}_{-0.4} \times 10^{-2} & -12.183 \pm 0.007 & -12.32 \pm 0.04 & 8.12^{+0.57}_{-0.51} & 7.56^{+0.43}_{-0.56} & 0.31 \pm 0.08 & 0.38 \pm 0.02 & 2056.9  \\
12 & (2.6 \pm 1.4) \times 10^{-2} & (4.4 \pm 0.3) \times 10^{-2} & -12.095 \pm 0.006 & -11.93^{+0.01}_{-0.02} & 7.77^{+0.49}_{-0.45} & 7.99^{+0.41}_{-0.49} & 0.26^{+0.07}_{-0.06} & 0.29 \pm 0.02 & 2172.3  \\
13 & 6.7^{+2.3}_{-2.0} \times 10^{-2} & 9.3^{+0.4}_{-0.5} & -12.306 \pm 0.008 & -14.31^{+0.62}_{-0.51} & 8.31^{+0.87}_{-0.67} & 13.67^{+1.05}_{-1.18} & 0.41^{+0.10}_{-0.09} & 2.12^{+0.20}_{-0.25} & 2003.8  \\
14 & (3.9 \pm 1.9) \times 10^{-2} & (3.4 \pm 0.3) \times 10^{-2} & -12.254^{+0.007}_{-0.008} & -12.21 \pm 0.03 & 9.35^{+0.93}_{-0.76} & 12.87^{+1.06}_{-1.17} & 0.26^{+0.09}_{-0.08} & 0.29 \pm 0.02 & 2113.5  \\
15 & 2.3^{+1.6}_{-1.4} \times 10^{-2} & (4.9 \pm 0.3) \times 10^{-3} & -12.195 \pm 0.007 & -12.29 \pm 0.03 & 8.45^{+0.65}_{-0.55} & 7.57^{+0.52}_{-0.43} & 0.38^{+0.09}_{-0.08} & 0.44 \pm 0.03 & 2096.2  \\
16 & 1.9^{+1.5}_{-1.2} \times 10^{-2} & (7.0 \pm 0.5) \times 10^{-3} & -12.113 \pm 0.006 & -12.22^{+0.02}_{-0.03} & 10.09^{+1.08}_{-0.79} & 13.79^{+0.63}_{-0.77} & 0.20^{+0.09}_{-0.08} & 0.020 \pm 0.001 & 2200.8 \\
17 & 2.0^{+1.3}_{-1.2} \times 10^{-2} & 3.2^{+0.3}_{-0.2} \times 10^{-2} & -11.958^{+0.006}_{-0.005} & -11.96 \pm 0.01 & 10.00^{+0.87}_{-0.65} & 11.25^{+0.70}_{-0.73} & 0.27 \pm 0.08 & 0.28 \pm 0.02 & 2133.1 \\
18 & (4.7 \pm 1.9) \times 10^{-2} & 9.1^{+0.5}_{-0.7} & -12.045^{+0.007}_{-0.006} & -14.58^{+0.56}_{-0.32} & 9.99^{+1.15}_{-0.90} & 5.64^{+0.44}_{-0.37} & 0.28 \pm 0.10 & 0.020 \pm 0.002 & 2413.7 \\
\enddata
\tablecomments{Results of spectral fits to the X-ray data with $1\sigma$ uncertainties. Spectral fitting was done two different ways: analysis of the projected annuli (under the ``Proj.'' columns) and a deprojected analysis (under the ``Deproj.'' columns) assuming that the elliptical annuli deprojected as if they were concentric.  The model parameters are the redshifted, intrinsic absorption column in units of $10^{22}\ \units{cm^2}$ ($N_{\mathrm{H}}$), the logarithmic 0.3--7 keV flux in units of \units{erg\ s^{-1}\ cm^{-2}} ($\log{F_X}$), the temperature of the collisional plasma of the APEC model in units of keV ($kT$), and the abundance relative to solar ($Z/Z\subsun$).  The final column lists $C$-stat values per 2097 degrees of freedom for each annulus for the best fits of the projected models.  The deprojected fit, which was fitted to all annuli simultaneously, had best-fit $C$-stat of 35441.6 for 37800 degrees of freedom. In the projected case, the annuli fit are independent of each other, but for the deprojected case, all parameter uncertainties are correlated.  The statistical uncertainties shown in the table come from the MCMC runs and are marginalized over all other parameters.  In addition to the statistical uncertainties, there are systematic uncertainties to the deprojected fits arising from using a concentric model for non-concentric data.  Thus the full uncertainty is larger than what is shown and likely a large effect for the scatter seen in the profiles of Figure \ref{fig:profiles}.}
\label{tab:specfit}
\end{deluxetable*}

\end{rotatetable*}

\begin{figure*}[hbt!]
\includegraphics[width=0.49\textwidth]{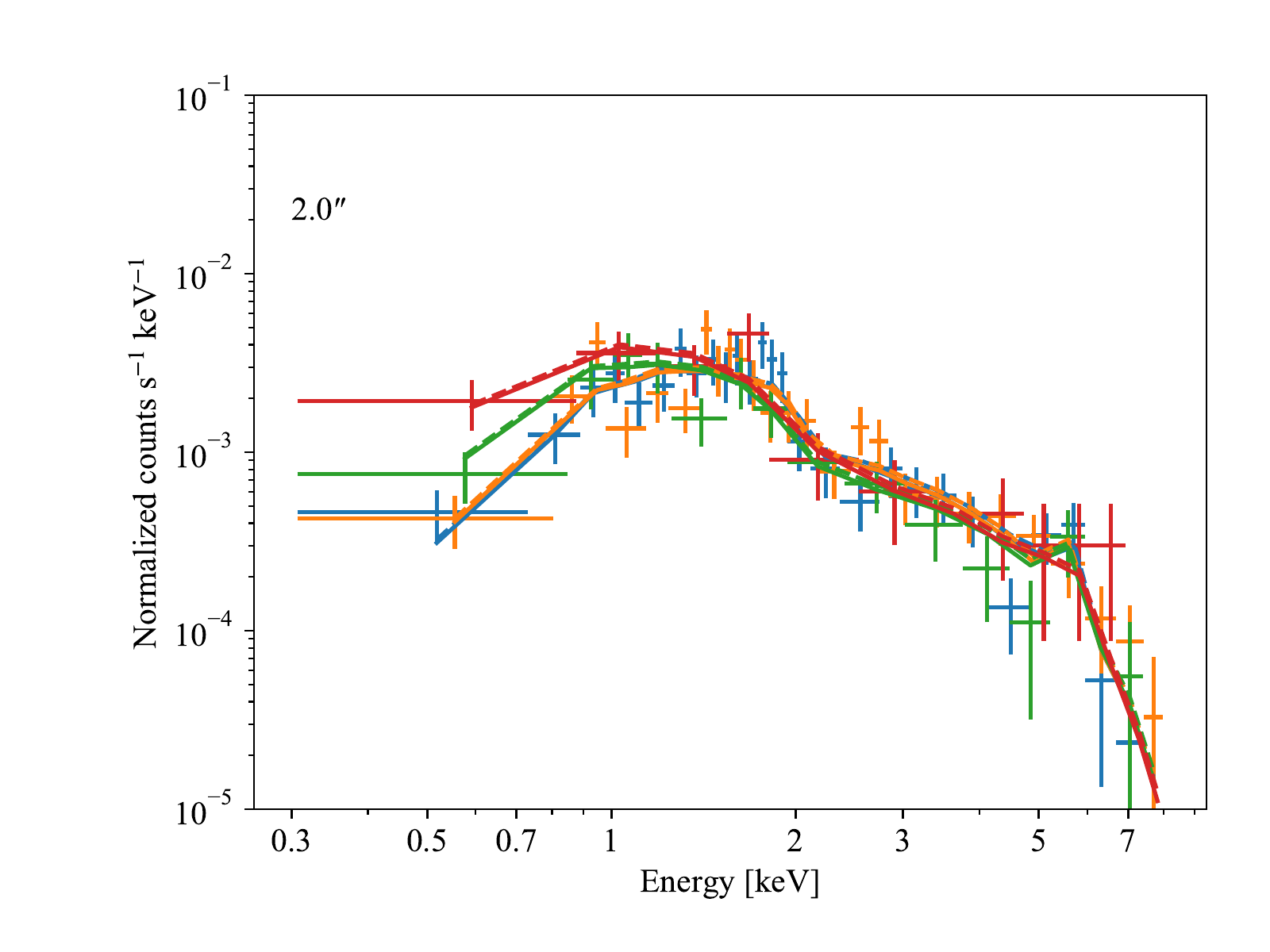}
\includegraphics[width=0.49\textwidth]{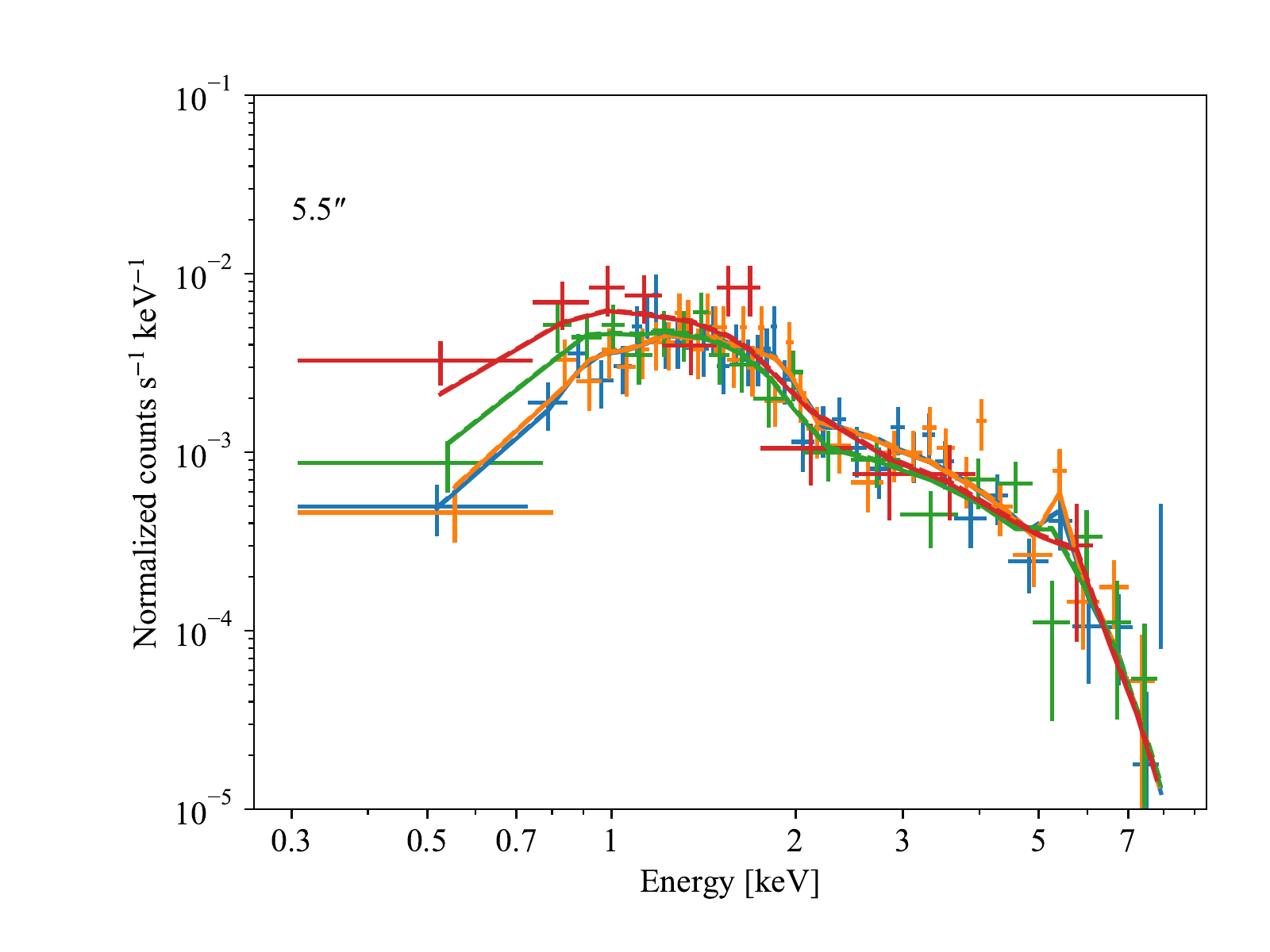}\\
\includegraphics[width=0.49\textwidth]{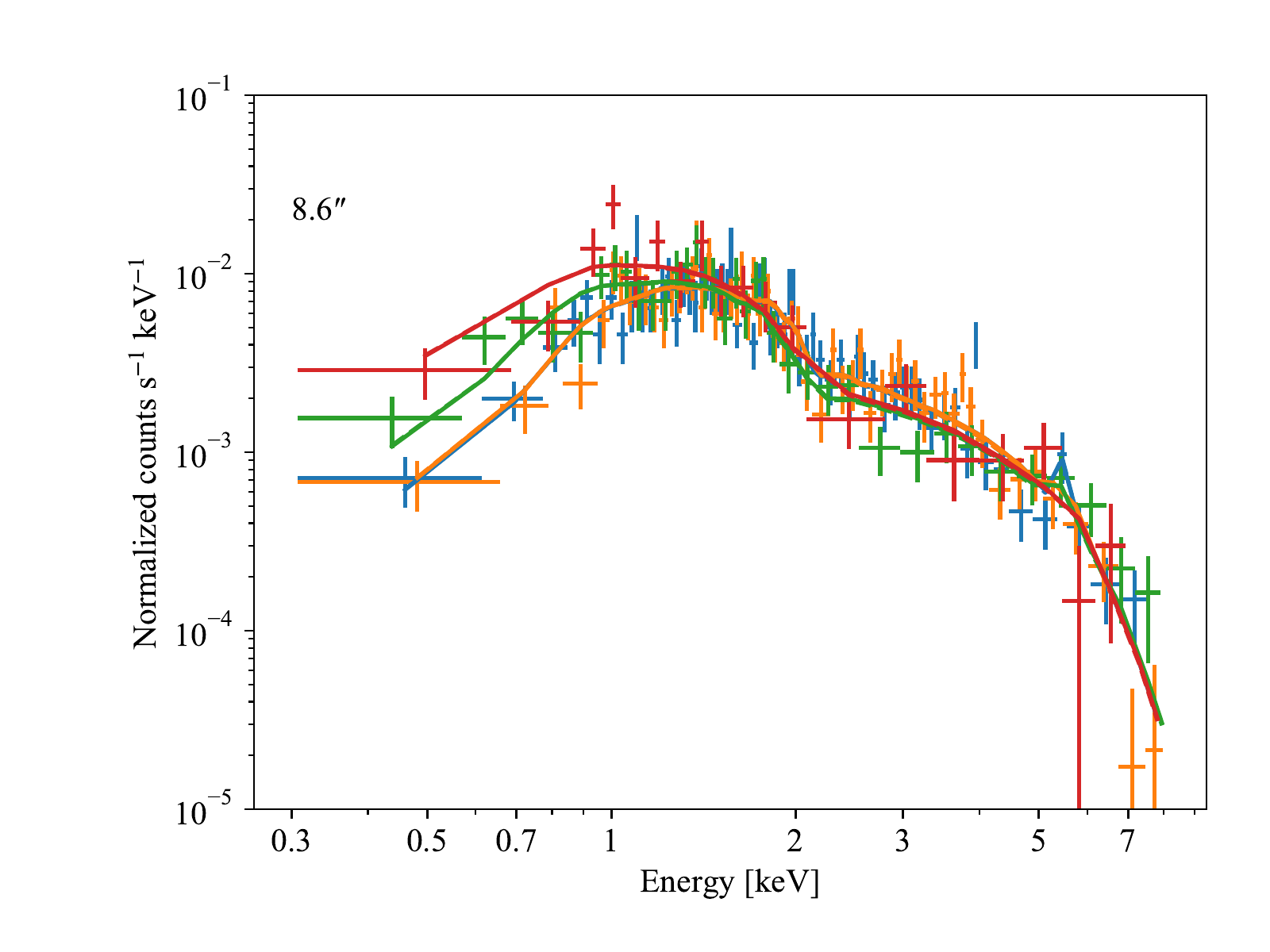}
\includegraphics[width=0.49\textwidth]{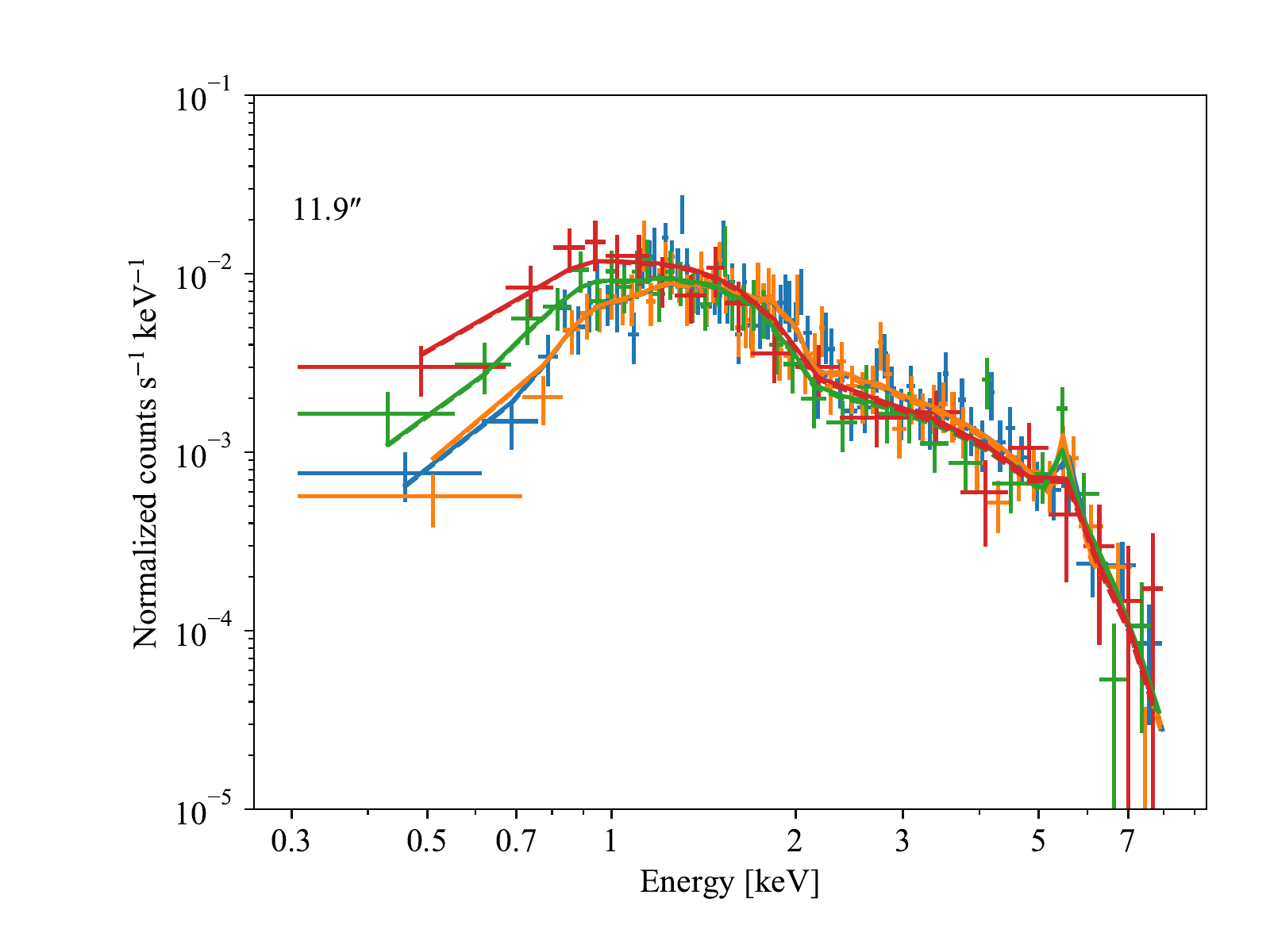}\\
\includegraphics[width=0.49\textwidth]{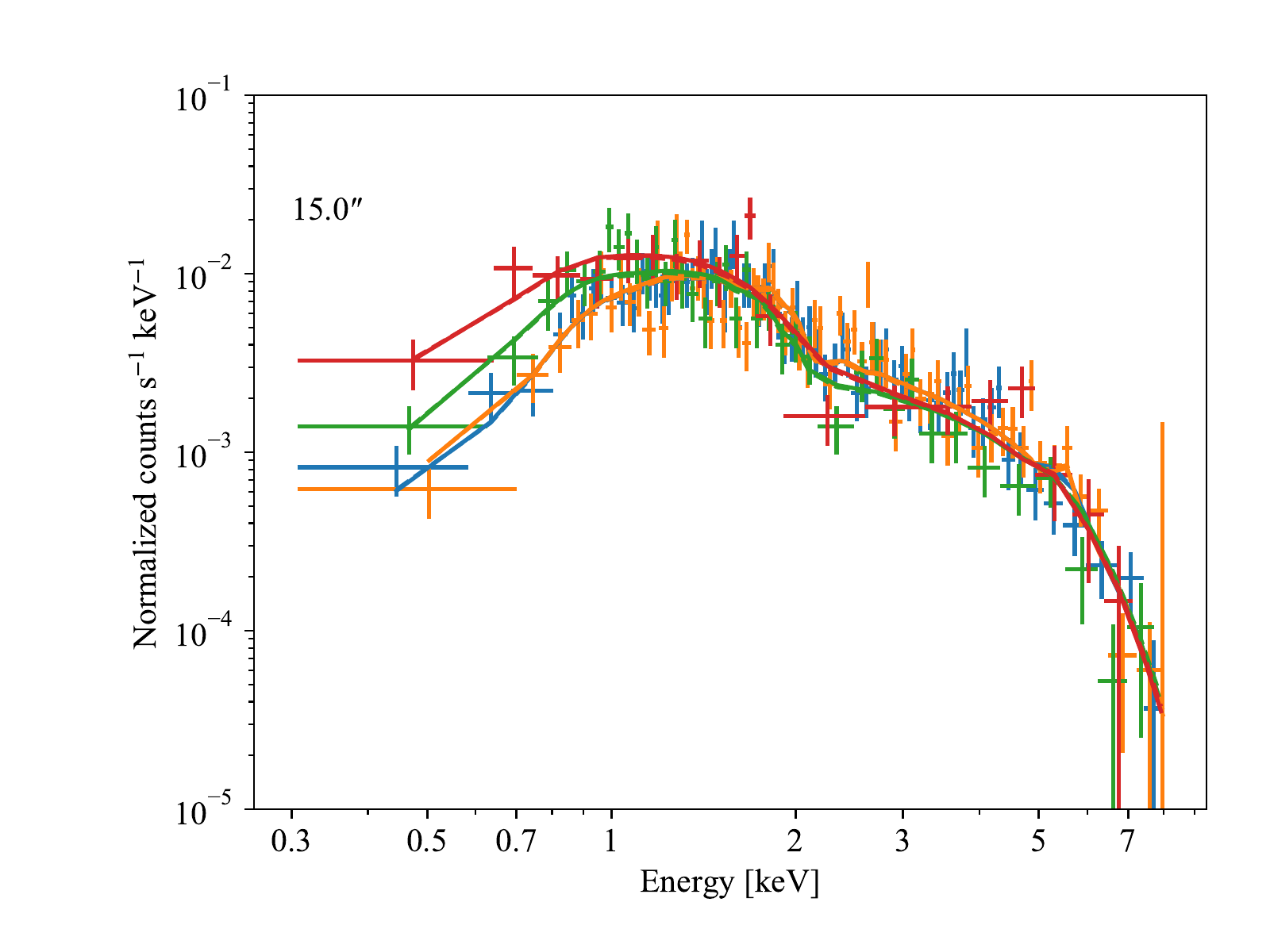}
\includegraphics[width=0.49\textwidth]{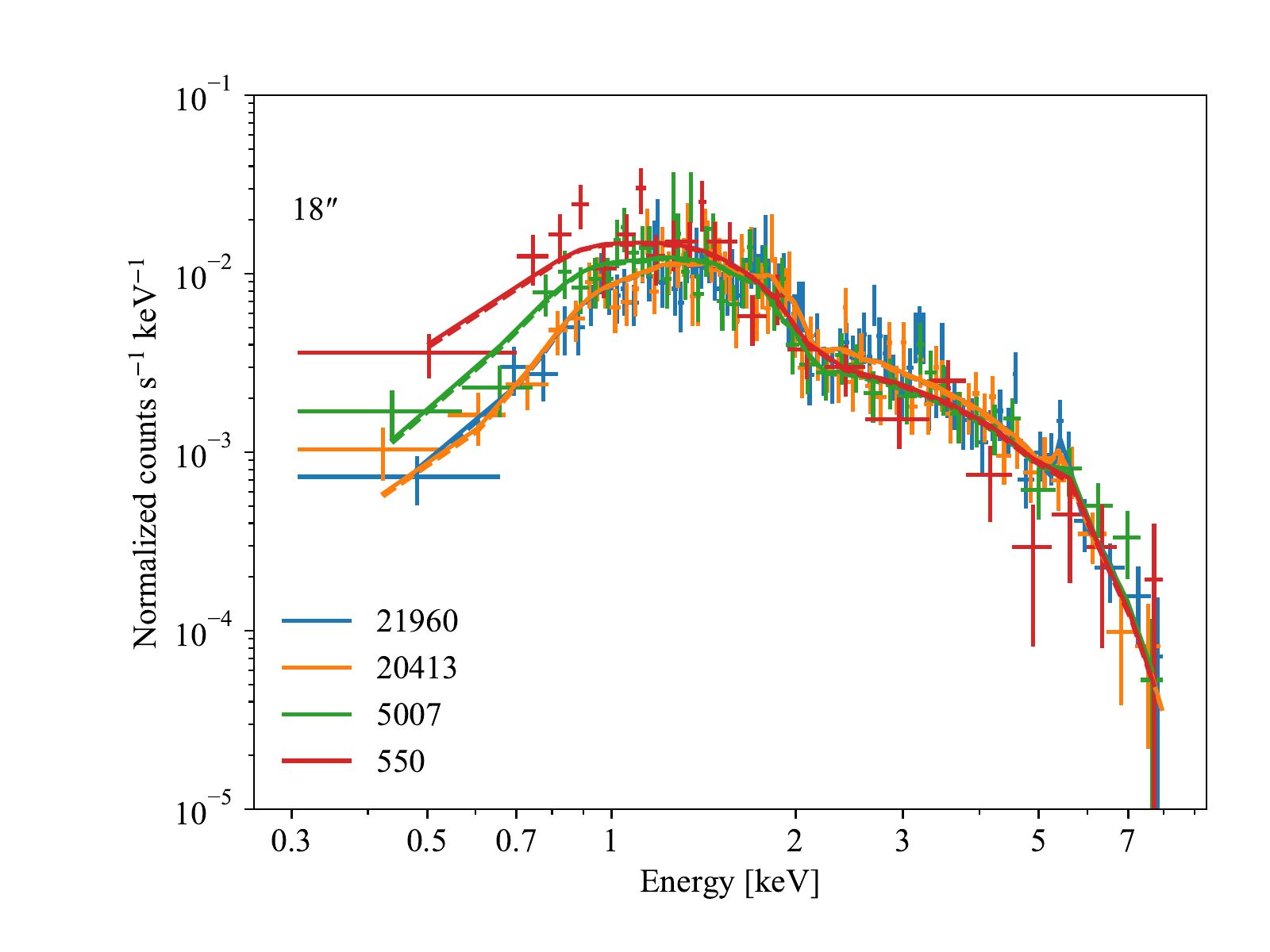}\\
\caption{Spectra and fits for each of the inner six annuli.  Spectra are the \emph{Chandra} spectra of the ObsID indicated by the color in the legend.   The corresponding lines are the best-fit models folded through the instrument response with solid line as the results from the fits to the projected annuli and the dashed lines are results from the deprojected models.  The model parameters for each ObsID are identical and only differ because of the changing response.  The label in the top-left corner indicates the average radius of the annulus.
\label{fig:profilesa}}
\end{figure*}

\begin{figure*}[ht!]
\includegraphics[width=0.49\textwidth]{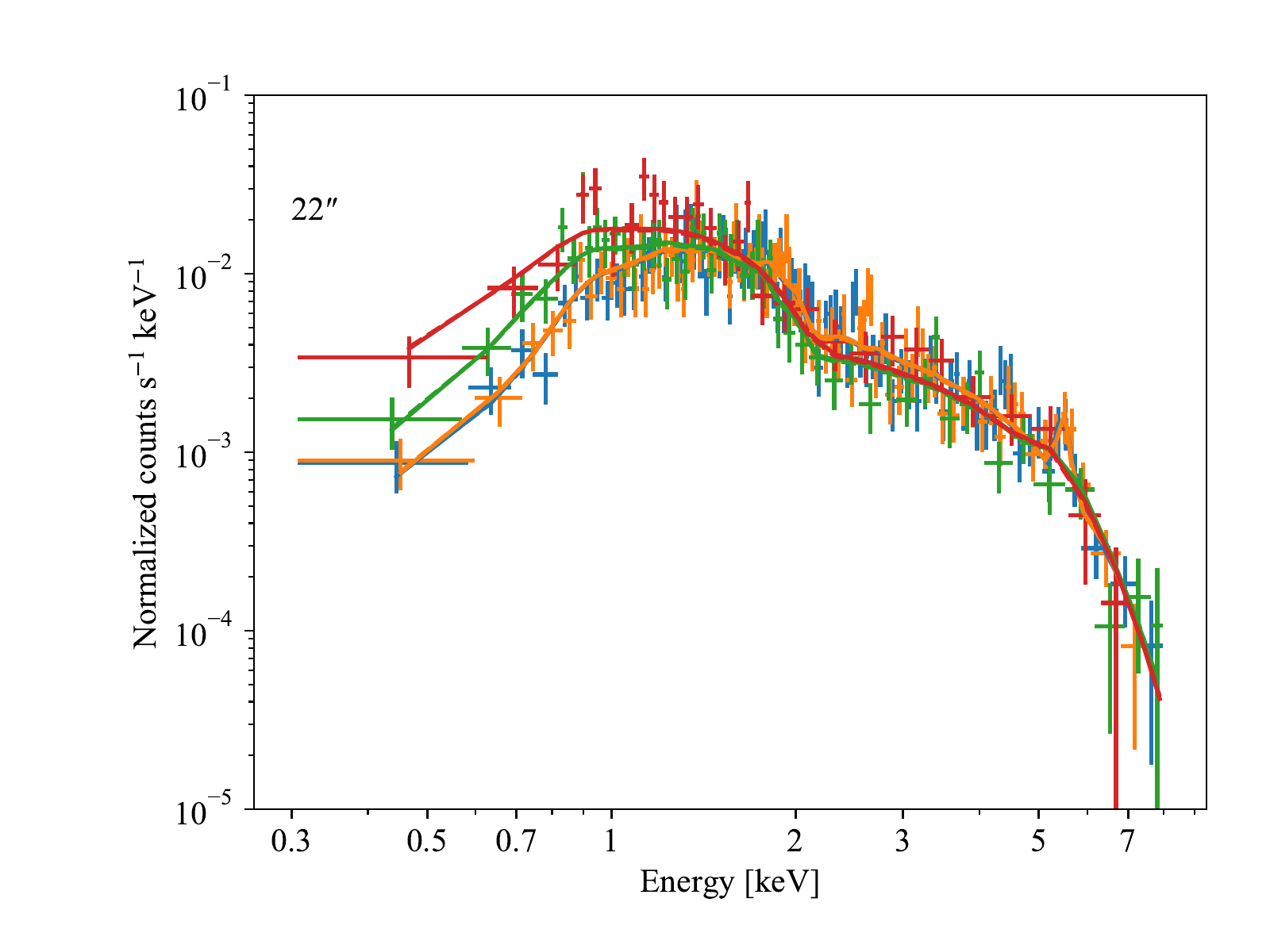}
\includegraphics[width=0.49\textwidth]{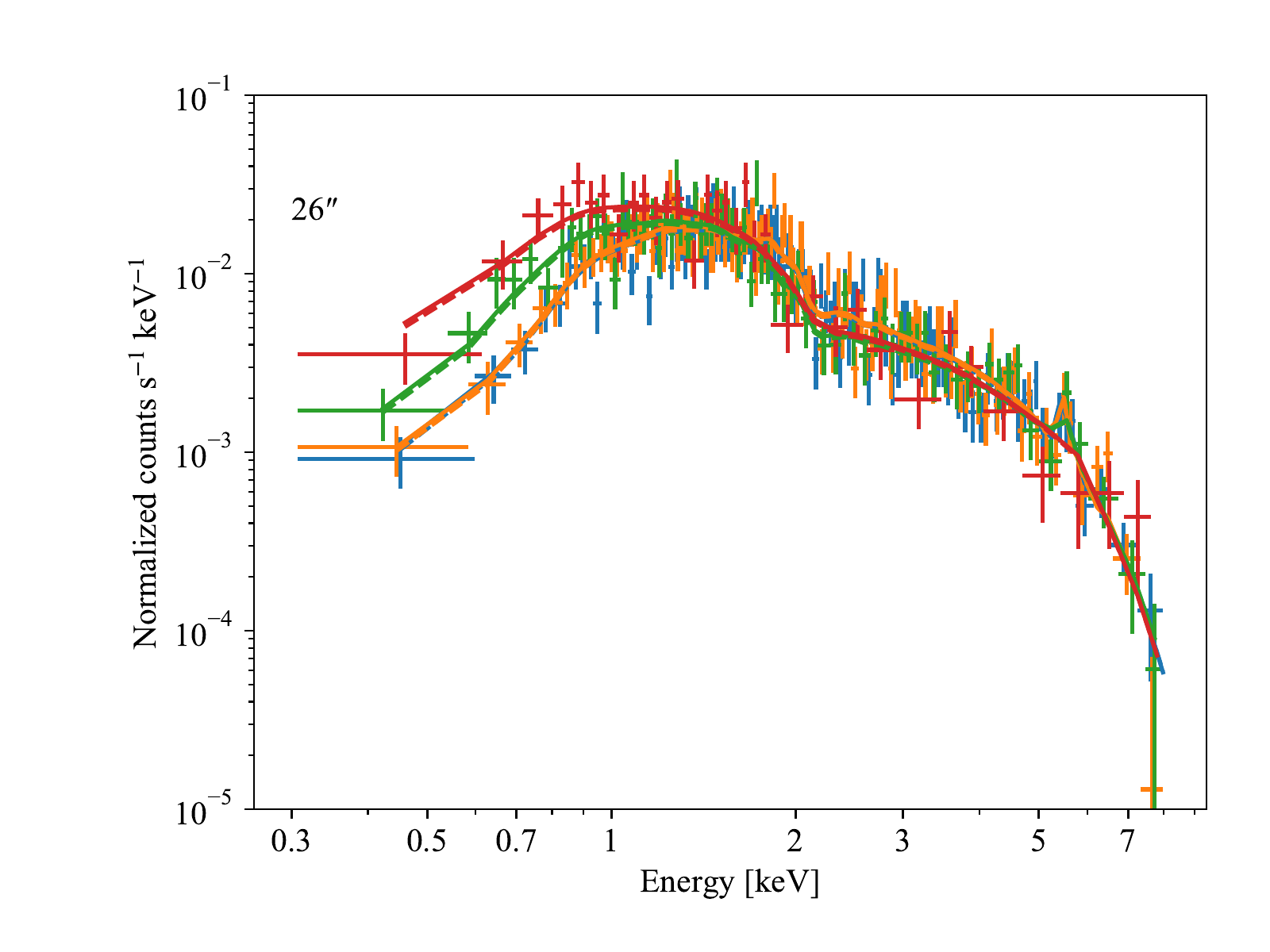}\\
\includegraphics[width=0.49\textwidth]{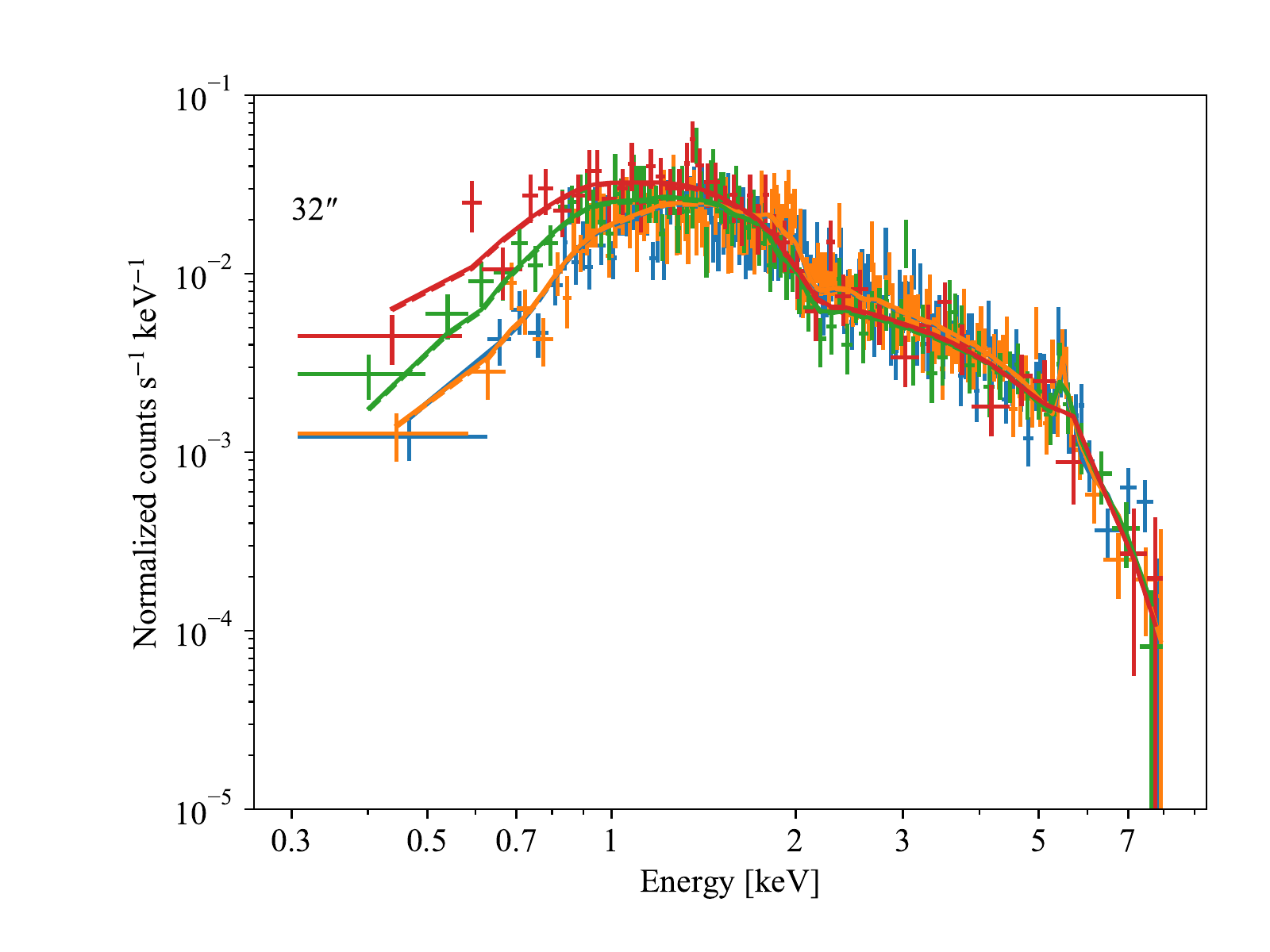}
\includegraphics[width=0.49\textwidth]{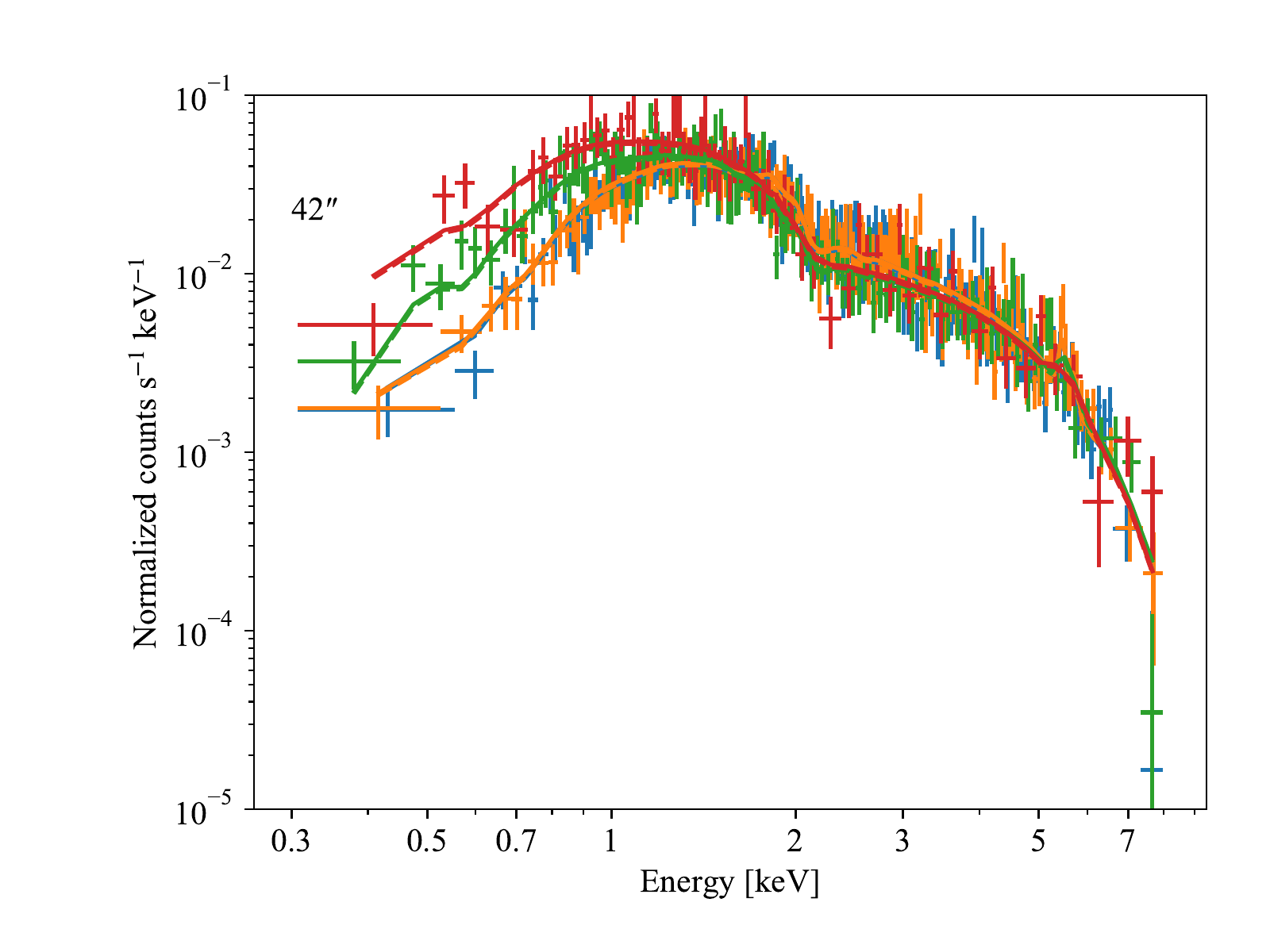}\\
\includegraphics[width=0.49\textwidth]{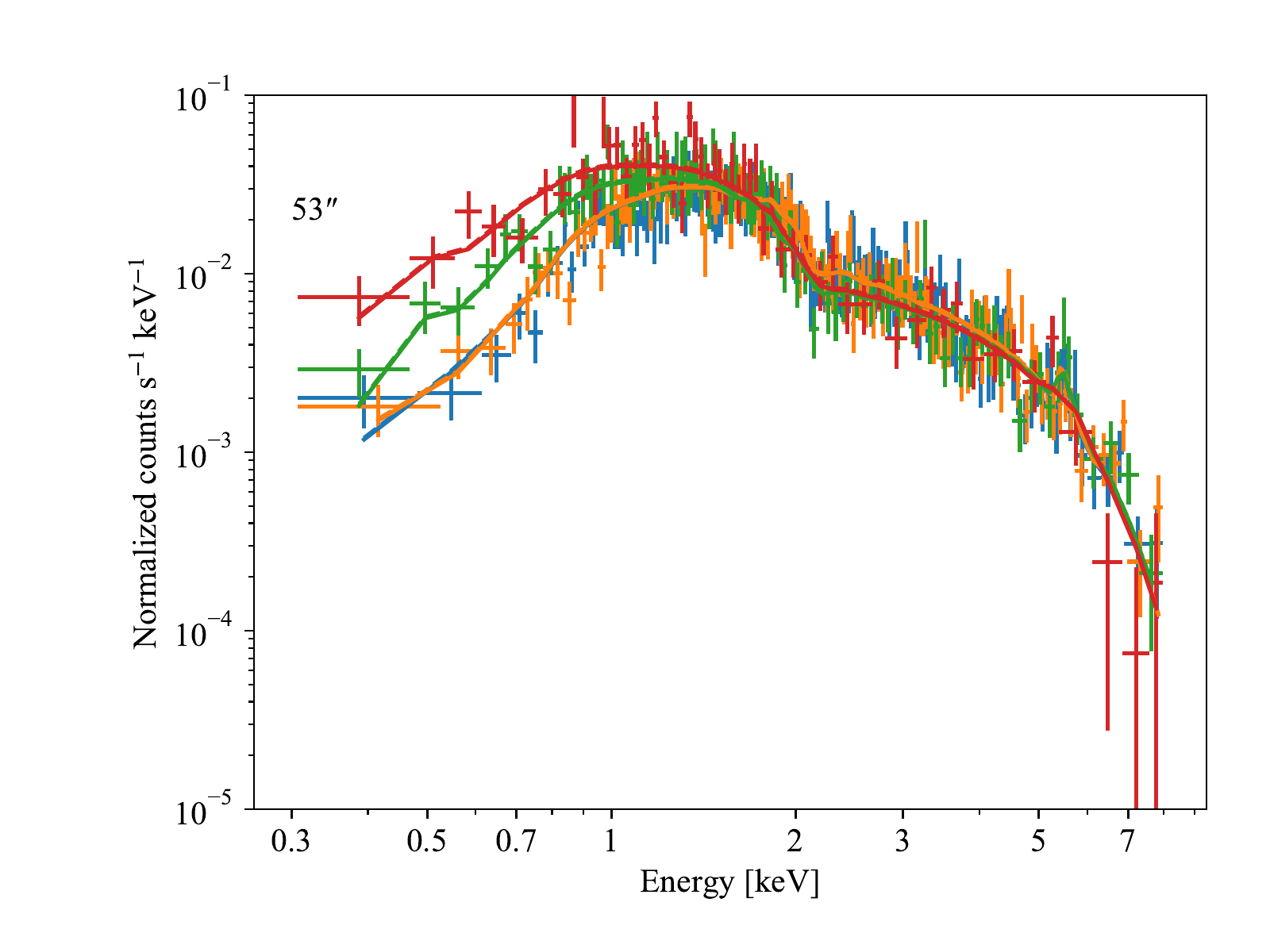}
\includegraphics[width=0.49\textwidth]{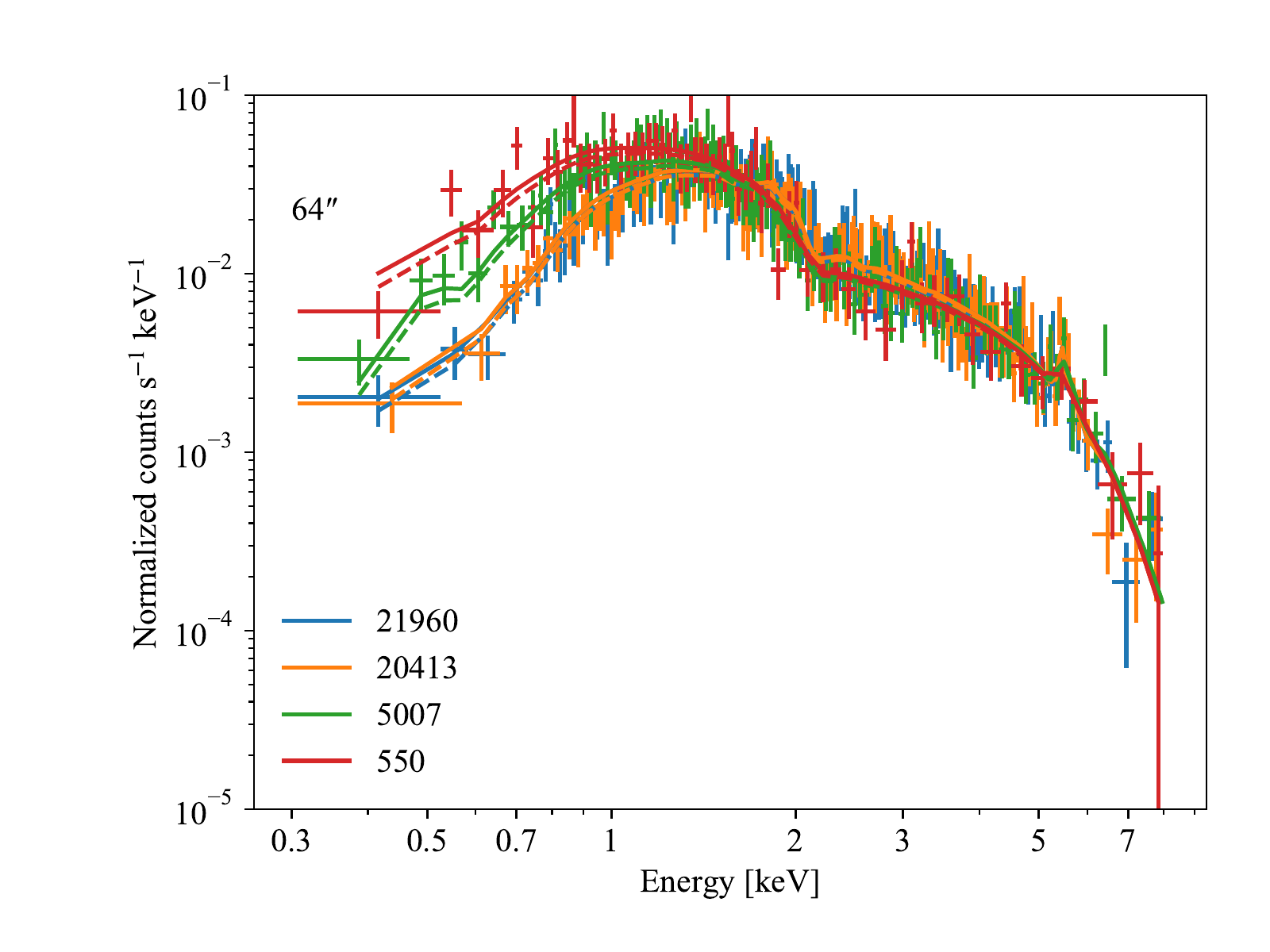}\\
\caption{Same as Fig.\ \ref{fig:profilesa} but for the central six annuli.
\label{fig:profilesb}}
\end{figure*}

\begin{figure*}[ht!]
\includegraphics[width=0.49\textwidth]{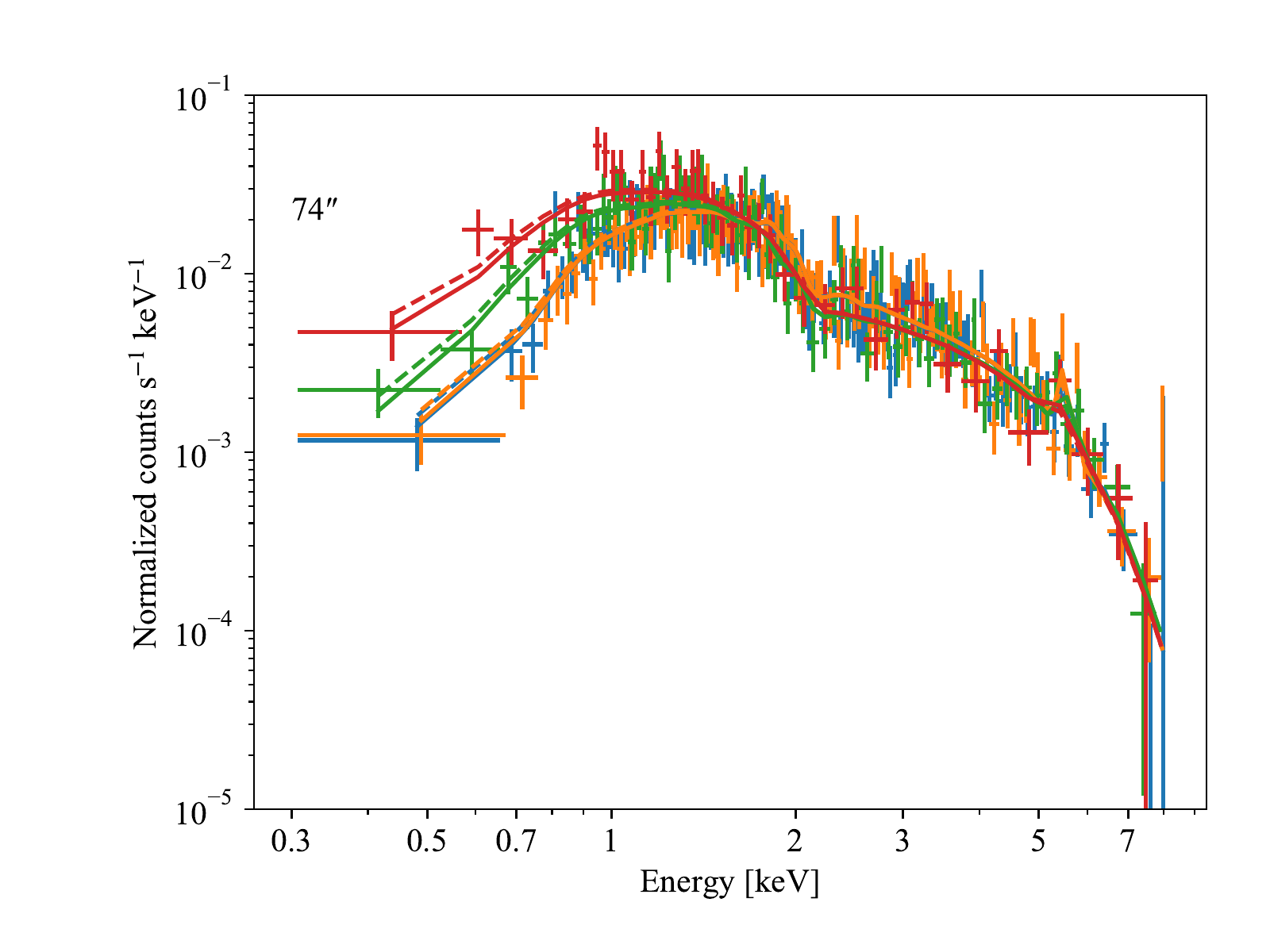}
\includegraphics[width=0.49\textwidth]{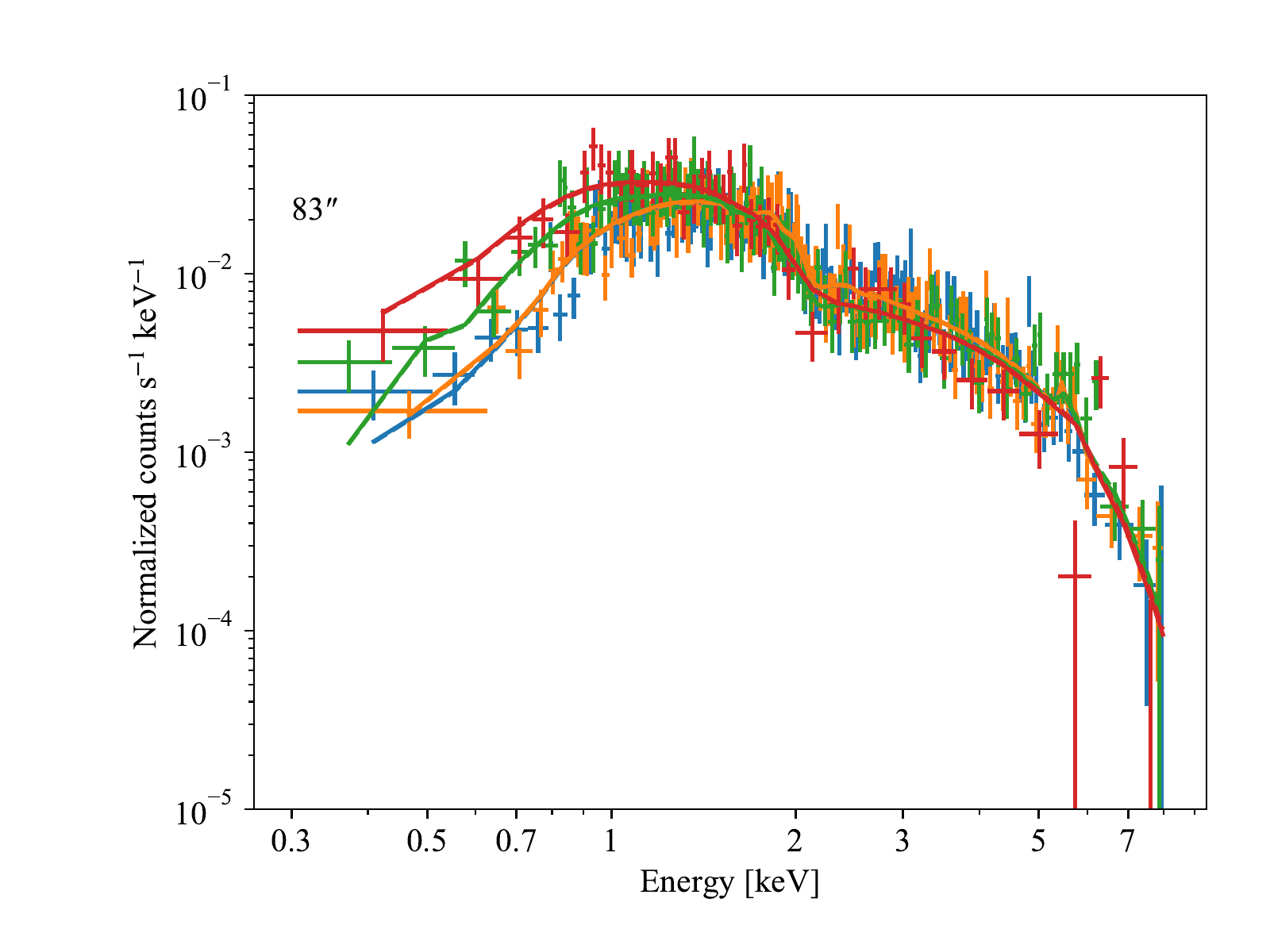}\\
\includegraphics[width=0.49\textwidth]{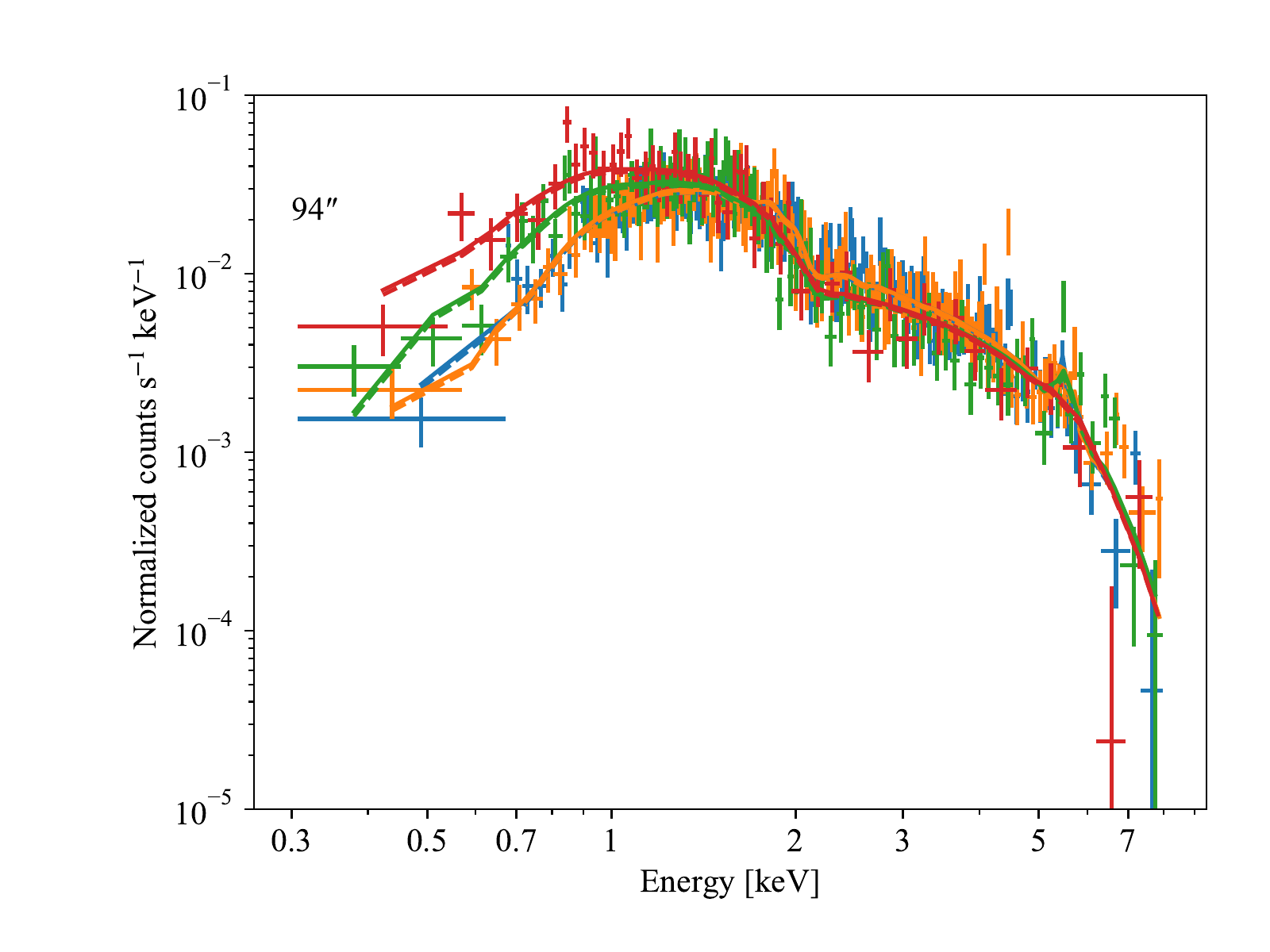}
\includegraphics[width=0.49\textwidth]{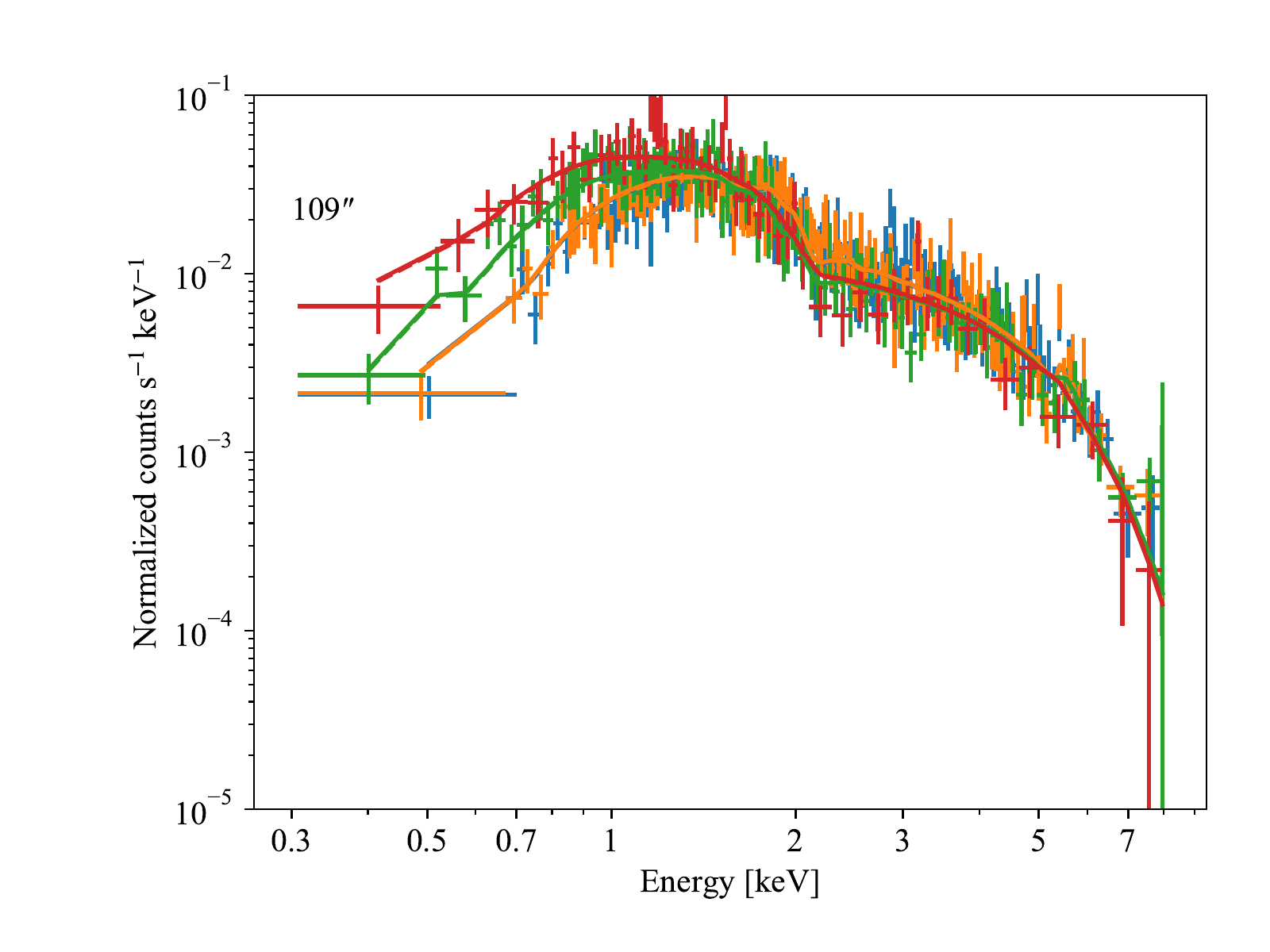}\\
\includegraphics[width=0.49\textwidth]{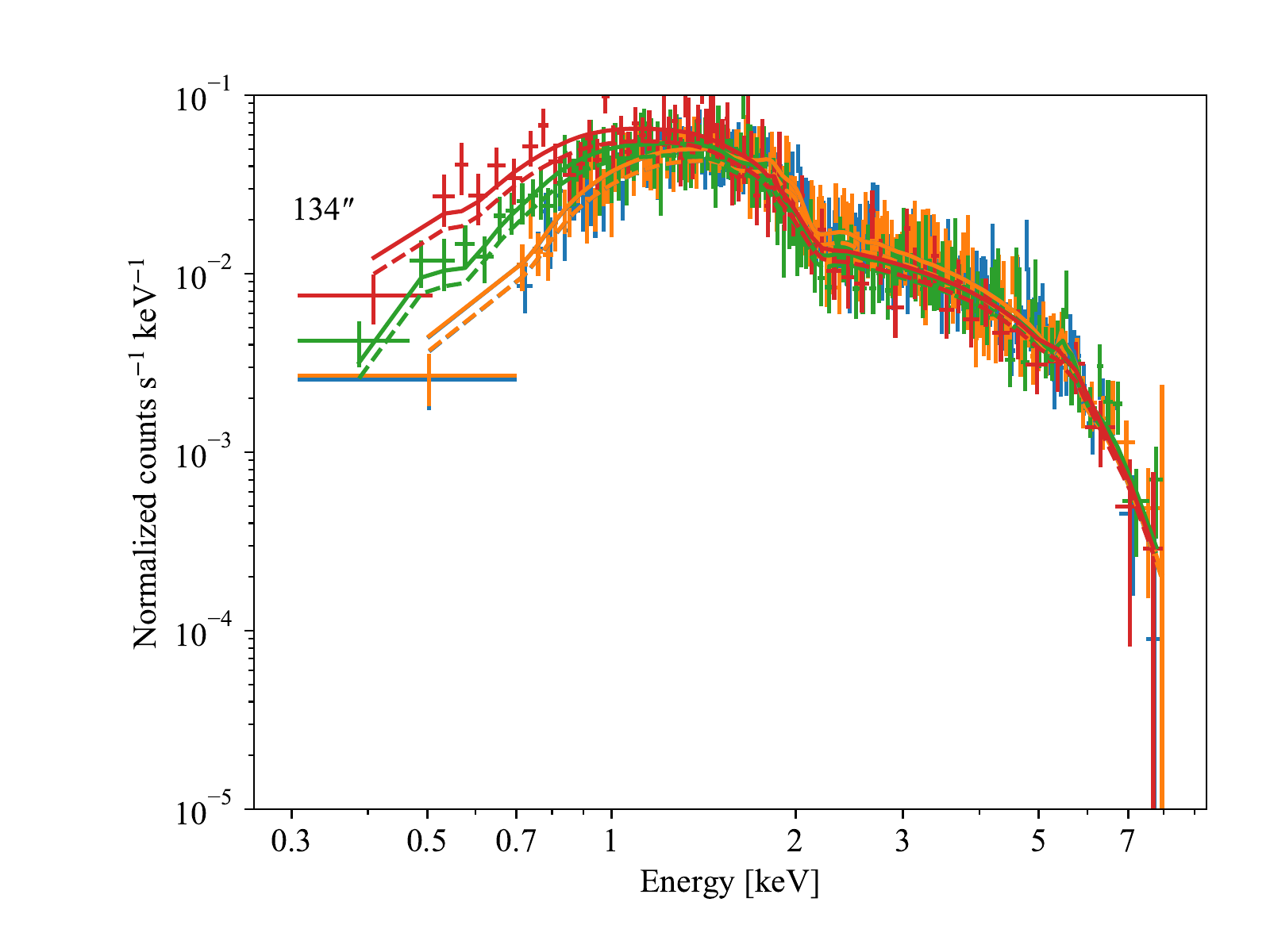}
\includegraphics[width=0.49\textwidth]{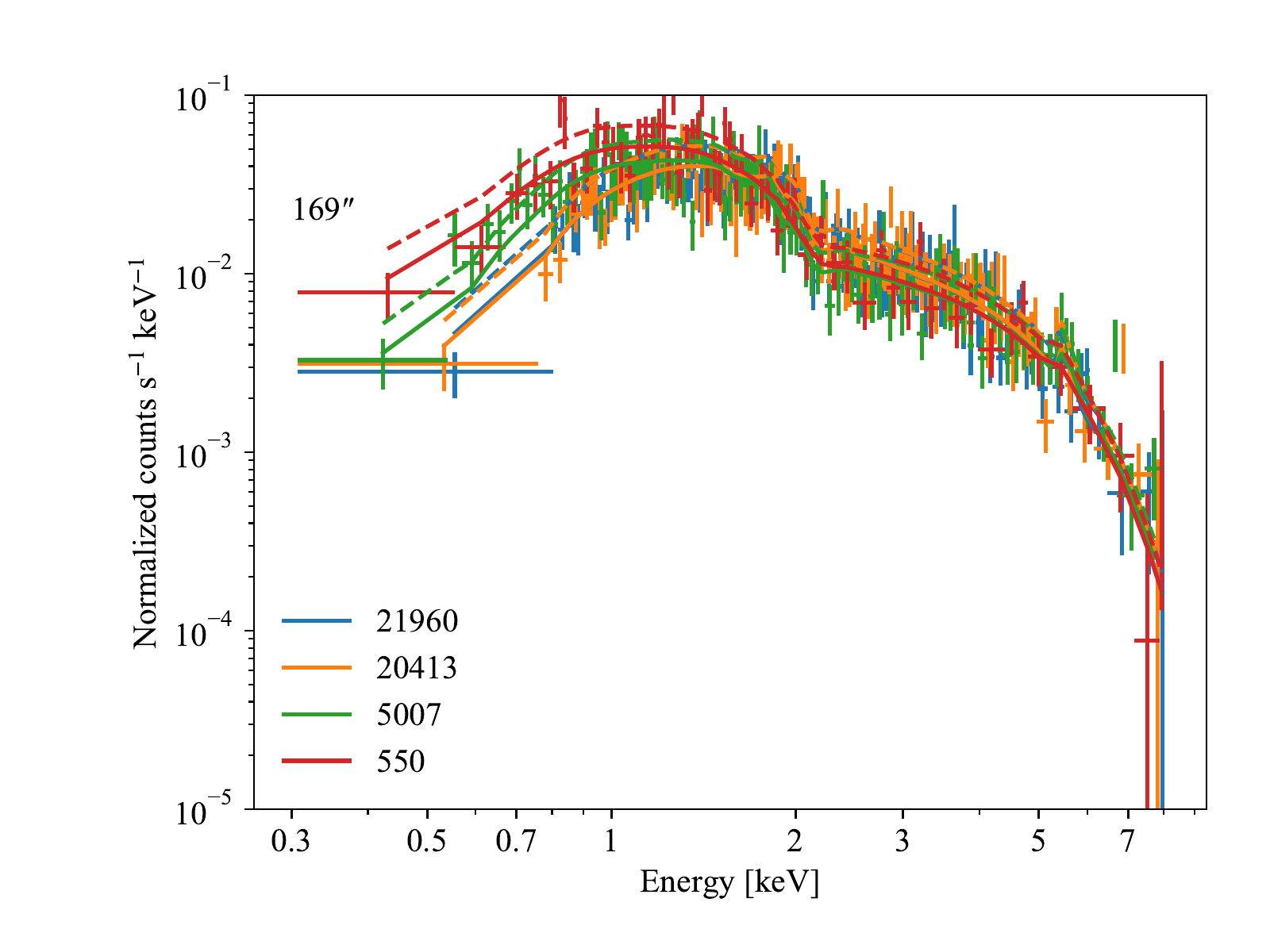}\\
\caption{Same as Fig.\ \ref{fig:profilesa} but for the outer six annuli.
\label{fig:profilesc}}
\end{figure*}

\end{document}